\documentclass[]{IEEEtran}
\usepackage{amsmath,amsthm,amsfonts,amssymb}
\usepackage{cite}
\usepackage{graphicx}
\usepackage{bbm}
\usepackage{braket}
\usepackage{textcomp,gensymb}
\usepackage{xcolor}
\usepackage[hidelinks]{hyperref}
\usepackage{needspace}
\newcommand{\eref}[1]{(\ref{#1})}

\newtheorem{theorem}{Theorem}
\newtheorem{corollary}{Corollary}
\newtheorem{lemma}{Lemma}
\newtheorem{remark}{Remark}
\newtheorem{assumption}{Assumption}

\DeclareMathOperator{\ind}{\mathbbm{1}}
\newcommand*\diff{\mathop{}\!\mathrm{d}}
\DeclareMathOperator*{\argmin}{arg\,min}

\title{From Footprints to Service Regions: Modeling and Analysis of Dense LEO Satellite Networks Using an Orbit-Aware Cox Point Process}
\author{Chang-Sik~Choi,~\IEEEmembership{Member,~IEEE}
\thanks{This work has been accepted for publication in the IEEE Transactions on Wireless Communications. \copyright~2026 IEEE. Personal use of this material is permitted. Permission from IEEE must be obtained for all other uses, in any current or future media, including reprinting/republishing this material for advertising or promotional purposes, creating new collective works, for resale or redistribution to servers or lists, or reuse of any copyrighted component of this work in other works.}
\thanks{The author is with the School of Electrical Engineering, KAIST, Daejeon, South Korea (e-mail: changsik@kaist.ac.kr).}
}

\begin{document}
\maketitle

\begin{abstract}
In dense low Earth orbit (LEO) satellite networks, several accessible satellites cover the same location. A footprint therefore identifies where communication is feasible but does not determine which satellite serves each user. The service region of a satellite, namely the set of locations it serves, governs satellite load, resource sharing, and user throughput. Using an orbit-aware Cox point process on a common orbital shell, we model each service region as the intersection of a satellite's spherical-cap footprint and its spherical Voronoi cell under nearest-satellite association. We then use Palm calculus to derive exact expressions for the mean area and perimeter and a coverage--load identity. After coverage saturates, mean area and load decrease inversely with population. In contrast, adding satellites along existing orbits leaves a positive limiting mean perimeter. We connect this geometry to an approximate noise-limited user rate. Along the tested densification paths, the increase in user rate comes mainly from the decrease in mean satellite load rather than from the improvement of the serving link. We also compare the model with Walker constellations with Starlink-type parameters, using Cox parameters matched to the local orbit and satellite densities at each latitude. The model reproduces the mean area and perimeter of these constellations over broad latitude ranges, with larger differences near component maximum latitudes. These results show that service-region geometry, and hence satellite load and user rate, depend not only on the total satellite population but also on its allocation across orbital planes.
\end{abstract}

\begin{IEEEkeywords}
Cox point processes, load-aware rate, low-Earth-orbit satellite networks, service regions, spherical Voronoi tessellations, stochastic geometry.
\end{IEEEkeywords}

\section{Introduction}
\label{sec:introduction}
\subsection{Motivation}
\label{subsec:motivation}
Large low Earth orbit (LEO) constellations are being deployed to provide broadband access, backhaul, Internet of Things connectivity, and emergency communications~\cite{3gppNTNOverview,azari2022evolution}. As these constellations grow to thousands of satellites, a ground user typically lies within the footprints of several satellites at the same time~\cite{starlinkFCC,kuiperFiling,simulationStudies}. In this regime, the main question is no longer whether a user is covered. The network now needs to determine which accessible satellite serves the user and how many other users share the resources of that satellite.

Both quantities are determined by the region that each satellite \emph{serves}. The footprint of a satellite is the set of locations at which it is accessible, whereas its spherical Voronoi cell is the set of locations for which it is the nearest satellite. Under nearest-satellite association on a common altitude shell, the service region of a satellite is the intersection of these two sets. In terms of network performance, the area of the service region determines the mean number of users that the satellite serves under homogeneous traffic. Its boundary consists of Voronoi edges, where the serving satellite changes, and of the retained footprint rim, where coverage ends. Each Voronoi edge lies on the bisector between the satellite and one competitor. That competitor is either a same-orbit neighbor, whose position is set by the spacing along the shared orbit, or an other-orbit satellite, whose position is set by the arrangement of the other orbital planes. Which group supplies the nearer competitors determines the shape of the service region.

Service-region statistics are also critical when satellites are shared among operators or partitioned into network slices, because each service then relies on a subset of the accessible satellites~\cite{ntnSharing,sliceNTN,sharingTerrestrial}. For constellation design, the key question is how these statistics change with the number of orbital planes and the number of satellites per orbit. An analytical characterization of service regions exposes this dependence without a simulation campaign for each configuration. It describes spatial load and association boundaries at a given time. Once satellite motion and an association policy are accounted for, it can also describe handover rates.

\subsection{Related Work}
\label{subsec:related_work}
Answering this question requires a model of satellite positions that retains the orbital structure of a constellation. Deployed constellations follow Walker configurations, which prescribe the number of orbital planes, the number of satellites per plane, and their relative phases~\cite{walkerClassic}. For a specified configuration, system simulations evaluate throughput, latency, and cost~\cite{simulationStudies}. However, characterizing the dependence of service-region statistics on orbital allocation requires repeated evaluations across configurations.

Stochastic geometry expresses this dependence analytically by treating satellite positions as a point process on the spherical shell~\cite{daley2007introduction,baccelli2010stochastic}. It has been widely used to analyze coverage and rate in terrestrial cellular networks~\cite{andrews2011tractable,dhillon2012modeling}. For satellite networks, binomial and Poisson point processes on the sphere yield downlink coverage and rate expressions of the typical user~\cite{okati2020downlink}, gateway-assisted coverage across altitudes~\cite{talgat2020stochastic}, and tractable visibility and nearest-distance distributions~\cite{al2021analytic,alhourani2021altitude}. Later work represented latitude-dependent nonuniform deployments~\cite{lee2024walker} and incorporated shadowed-Rician fading into the link analysis~\cite{ppp_extensions}. These models place satellites independently and therefore do not represent the spatial dependence induced by orbital planes, which is the structure that distinguishes same-orbit from other-orbit competitors.

The orbit-aware Cox process of~\cite{choi2022analytical} retains this spatial structure by generating orbital planes first and then placing satellites conditionally on those planes. Using this model, \cite{choi2022analytical} analyzed visibility, contact distance, and downlink coverage. The multi-altitude construction in~\cite{choi2024cox} and the heterogeneous-network analysis in~\cite{choi2024exploiting} extend the model to further deployment settings, and the latter compares its coverage predictions with Walker constellations. These results characterize coverage and the serving link. They do not characterize the region that each satellite serves after association.

Several related concurrent works analyze other aspects of satellite networks. The infrastructure-sharing analysis in~\cite{choi2026sharing} evaluates footprint overlap and connectivity. The uplink analysis in~\cite{choi2026uplink} characterizes interference and SINR at a typical satellite. The coordination analysis in~\cite{choi2026coordination} addresses interference suppression and signaling overhead. The Walker analysis in~\cite{choi2026walker} studies the dynamics of deterministic constellations. None of them analyzes the association-induced partition of the covered surface or its area, perimeter, and load statistics.

Because service regions are footprint-constrained Voronoi cells, their area and perimeter statistics are closely related to those of a Voronoi tessellation. In the planar case, the moments of the typical-cell area and perimeter were derived for the Poisson--Voronoi tessellation~\cite{planarVoronoi}, and its area distribution was approximated by fitted Gamma distributions~\cite{gammaApprox,pineda2004cell}. On the sphere, mean characteristics and typical-cell distributions are known for uniformly distributed point processes~\cite{sphericalVoronoi,kabluchko2021typical}. Satellite studies use Voronoi cells in stochastic-model comparisons and coverage-oriented constellation design~\cite{okati2020downlink,jeon2025design}. Nevertheless, none of these results gives the footprint-constrained Voronoi cell statistics of an orbit-aware Cox process, in which same-orbit and other-orbit competitors produce different Voronoi boundaries around the typical cell. In this paper, we use the orbit-aware Cox model to characterize the service region, namely the intersection of a satellite's footprint and its spherical Voronoi cell. Through this characterization, we analyze the mean service-region area, the mean service-region perimeter, the satellite load, and the load-aware user rate, and we validate the analysis through simulations of both the Cox model and Walker constellations. To the best of our knowledge, no prior work provides these results for an orbit-constrained satellite point process.
\subsection{Contributions}
\label{subsec:contributions}
Our contributions are as follows.
\begin{itemize}
\item \textbf{Exact service-region area and mean satellite load.}
We separate same-orbit and other-orbit competitors under the Palm distribution to derive the retention probability and the exact mean normalized area. A coverage--load identity connects this geometry to mean satellite load without requiring the full cell-area distribution. Once coverage saturates, both mean area and mean load decrease inversely with population. This relation provides a reference for sizing per-satellite capacity to meet anticipated traffic demand.

\item \textbf{Service-boundary geometry and orbital allocation.}
We derive the exact mean normalized perimeter by separating the retained footprint rim from same-orbit and other-orbit Voronoi edges. With a fixed mean number of orbital planes, the perimeter approaches a positive limit as the along-orbit density grows, even though the mean area vanishes. Thus, load reduction does not imply uniform contraction of service regions. Orbital allocation determines their boundary geometry.

\item \textbf{Load-aware rate and densification gains.}
We combine the serving-distance distribution conditioned on coverage with an area-biased Gamma approximation and a link--load factorization to derive an approximate mean user rate. We assess the two approximations separately. In the tested regimes, load reduction contributes substantially more to the rate gains from densification than serving-link improvement. These rate gains diminish at light load. Once association limits the service region, additional footprint margin gives little rate benefit. Adding satellites along existing orbits does not reduce the distance to the nearest orbital plane. Orbital allocation therefore affects the rate at a fixed population. Gain and interference comparisons examine the load-reduction benefit beyond the noise-limited reference.

\item \textbf{Walker comparisons and applicability.}
We compare our predictions of mean area, perimeter, and rate with empirical means from a family of $53^\circ$ single-inclination Walker constellations and a five-component multi-inclination Walker constellation. Both comparisons use latitude-dependent matching of local orbit and satellite densities without fitting either service-region measure. The Cox model reproduces the mean area and perimeter of the Walker constellations over broad latitude ranges. This agreement supports the use of the analytical results for preliminary deployment comparisons. Larger differences occur near component maximum latitudes, where rapid density variation and directional structure require detailed geometric evaluation.
\end{itemize}

\Needspace{5\baselineskip}
\subsection{Organization}
Section~\ref{sec:system_model} defines the spatial, association, and traffic models. Section~\ref{sec:performance_analysis} derives the service-region and rate results and presents their numerical evaluation. Section~\ref{sec:conclusion} discusses the implications and extensions.

\section{System Model}
\label{sec:system_model}
\subsection{Spatial Model}
\label{subsec:spatial}

We consider a snapshot of a LEO satellite network deployed on a common spherical shell. Let $R_{\rm e}$ denote the Earth radius, $h$ the satellite altitude, and $R_{\rm s}=R_{\rm e}+h$ the shell radius. The Earth surface and satellite shell are $\mathbb S_{\rm e}=\{x\in\mathbb R^3:\|x\|=R_{\rm e}\}$ and $\mathbb S_{\rm s}=\{x\in\mathbb R^3:\|x\|=R_{\rm s}\}$, respectively.

We model satellite locations by an isotropic orbit-aware Cox point process on $\mathbb S_{\rm s}$, following the construction in~\cite{choi2022analytical,choi2024cox}. The model first generates random orbital planes and then places satellites conditionally on the resulting orbits, thereby preserving orbit-induced dependence while retaining analytical tractability.

We define the orbit process on the parameter space $\mathbb T=[0,\pi]\times[0,\pi]$, where each point $(\theta,\phi)\in\mathbb T$ represents an unoriented orbital plane, with boundary duplicates having zero intensity measure. The parameters $\theta$ and $\phi$ denote orbit longitude and inclination, respectively. We measure $\theta$ in the equatorial plane and $\phi$ between the orbit normal and the north-pole axis. Let $\Xi$ be a Poisson point process on $\mathbb T$ with intensity measure
\begin{equation}
\label{eq:orbit_intensity}
	\Lambda({\rm d}\theta,{\rm d}\phi)
	=
	\lambda\frac{\sin\phi}{2\pi}
	{\rm d}\theta{\rm d}\phi,
\end{equation}
where $\lambda$ is the mean number of orbital planes. Each point $(\theta_i,\phi_i)\in\Xi$ generates an orbit $\mathcal O_i=\mathcal O(\theta_i,\phi_i)$ on $\mathbb S_{\rm s}$.
Each orbit $\mathcal O(\theta,\phi)$ is a great circle of $\mathbb S_{\rm s}$, and we identify it with its \emph{pole}, the unit normal vector of the orbital plane,
\begin{equation}
	p_{\mathcal O}
	=
	(\sin\theta\sin\phi,\,-\cos\theta\sin\phi,\,\cos\phi).
	\label{eq:orbit_pole}
\end{equation}

Since the orbits are unoriented, $p$ and $-p$ represent the same orbit. A location $v$ is at angular distance $|\pi/2-\arccos(\braket{p,v/\|v\|})|$ from the orbit with pole $p$.
 Because $\sin\phi\,{\rm d}\theta\,{\rm d}\phi$ is the spherical surface element in the coordinates $(\theta,\phi)$, the orbit intensity measure $\Lambda$ maps to the uniform measure with total mass $\lambda$ on the sphere of unoriented poles, which expresses the isotropy of the orbit process.

Conditioned on $\Xi$, satellites are independently distributed on each orbit. On the orbit $\mathcal O_i$, the satellite phases $\{\omega_{i,j}\}_j$ form a one-dimensional Poisson point process on $[0,2\pi)$ with intensity $\mu/(2\pi)$, where $\mu$ is the mean number of satellites per orbit. Let $X_{i,j}$ denote the location of the $j$th satellite on the orbit $\mathcal O_i$, and let $\Phi_i=\sum_j\delta_{X_{i,j}}$ denote the satellite point process on that orbit. The overall satellite point process is
\begin{equation}
\label{eq:satellite_process}
	\Phi
	=
	\sum_i\Phi_i
	=
	\sum_i\sum_j\delta_{X_{i,j}}.
\end{equation}
Hence, the expected total number of satellites is $\mathbb E[|\Phi|]=\lambda\mu$, and the mean satellite density on $\mathbb S_{\rm s}$ is $\lambda\mu/(4\pi R_{\rm s}^2)$. The parameter $\lambda$ controls the number of orbital planes, whereas $\mu$ controls the satellite density along each orbit.

Owing to the rotation-invariant orbit-plane measure and uniform conditional placement of satellites along each orbit, $\Phi$ is isotropic. The satellite geometry observed from any Earth location is therefore statistically equivalent to that observed from any other location. Consequently, every performance measure determined by this geometry and location-independent random effects has the same distribution for all observers on $\mathbb S_{\rm e}$. Without loss of generality, we place the typical user at $u_0=(0,0,R_{\rm e})$.

All expectations below are ensemble averages over the random orbital planes and satellite phases. Uniform phase distributions remain uniform under deterministic phase shifts, so circular motion preserves the one-time distribution of the model. This stationarity does not establish ergodicity or identify ensemble averages with the time average of a fixed constellation. We do not model the temporal association process, Doppler, or handover decisions.

\subsection{Footprints and Service Regions}
\label{subsec:footprint}

We next distinguish a satellite's footprint from its association-induced service region. For $X\in\Phi$, let $\widehat X=X/R_{\rm s}$ denote its unit direction vector and $\bar X=R_{\rm e}\widehat X$ its subsatellite point on $\mathbb S_{\rm e}$. For an Earth location $u\in\mathbb S_{\rm e}$, define the Earth-centered angular separation from $X$ as $\gamma_X(u)=\arccos\!\left(u^\top X/(R_{\rm e}R_{\rm s})\right)$.

The footprint of satellite $X$ is modeled as a spherical cap centered at $\bar X$ with angular radius $\gamma_{\rm f}$:
\begin{equation}
\label{eq:footprint}
	\mathcal F_X
	=
	\{u\in\mathbb S_{\rm e}:\gamma_X(u)\leq\gamma_{\rm f}\}.
\end{equation}
The parameter $\gamma_{\rm f}$ summarizes the practical visibility constraints, including the minimum elevation angle and the antenna or beam coverage~\cite{3gppNTNOverview}. 
For a minimum elevation $\varepsilon_{\min}\geq0$, the geometric limit is $\gamma_{\rm elev}=\arccos((R_{\rm e}/R_{\rm s})\cos\varepsilon_{\min})-\varepsilon_{\min}$. Let $\gamma_{\rm snr}$ be the largest angular separation at which the mean received SNR exceeds a specified threshold $\mathrm{SNR}_{\min}$. If this SNR profile is nonincreasing, we use $\gamma_{\rm f}=\min\{\gamma_{\rm elev},\gamma_{\rm snr}\}$, with any additional beam limit included in the minimum. This defines accessibility by a mean link budget. Instantaneous fading or blockage can still cause an outage within the footprint. A nonmonotone or user-dependent accessibility region generally requires a different geometry.
 The footprint area and boundary length are $|\mathcal F_X|=2\pi R_{\rm e}^2(1-\cos\gamma_{\rm f})$ and $|\partial\mathcal F_X|=2\pi R_{\rm e}\sin\gamma_{\rm f}$, respectively. The union of all footprints, $	\mathcal B(\Phi)=\bigcup_{X\in\Phi}\mathcal F_X,$
defines the covered region. Since footprints may overlap, $\mathcal B(\Phi)$ is a spherical Boolean model and does not uniquely assign a serving satellite.

To define service regions, we consider nearest-accessible-satellite association. The spherical Voronoi cell generated by satellite $X$ is
\begin{equation}
\label{eq:voronoi_cell}
	\mathcal V_X(\Phi)
	=
	\{u\in\mathbb S_{\rm e}:
	\gamma_X(u)\leq\gamma_Y(u),\ \forall Y\in\Phi\}.
\end{equation}
Because all satellites lie on the common shell $\mathbb S_{\rm s}$, nearest-angular-distance association is equivalent to minimum-slant-range association. The collection $\{\mathcal V_X(\Phi):X\in\Phi\}$ therefore forms the ordinary spherical Voronoi tessellation generated by the satellite direction vectors. 
This association rule is a geometric reference. With equal transmit powers and a common nonincreasing mean gain profile, it also maximizes mean received power on the common shell.

The service region of satellite $X$ is the footprint-constrained Voronoi cell
\begin{equation}
\label{eq:service_region}
	\mathcal S_X(\Phi)
	=
	\mathcal F_X\cap\mathcal V_X(\Phi).
\end{equation}
The collection $\{\mathcal S_X(\Phi):X\in\Phi\}$ partitions the covered region $\mathcal B(\Phi)$ up to cell boundaries.

\begin{figure}[t]
	\centering
	\includegraphics[width=0.9\linewidth]{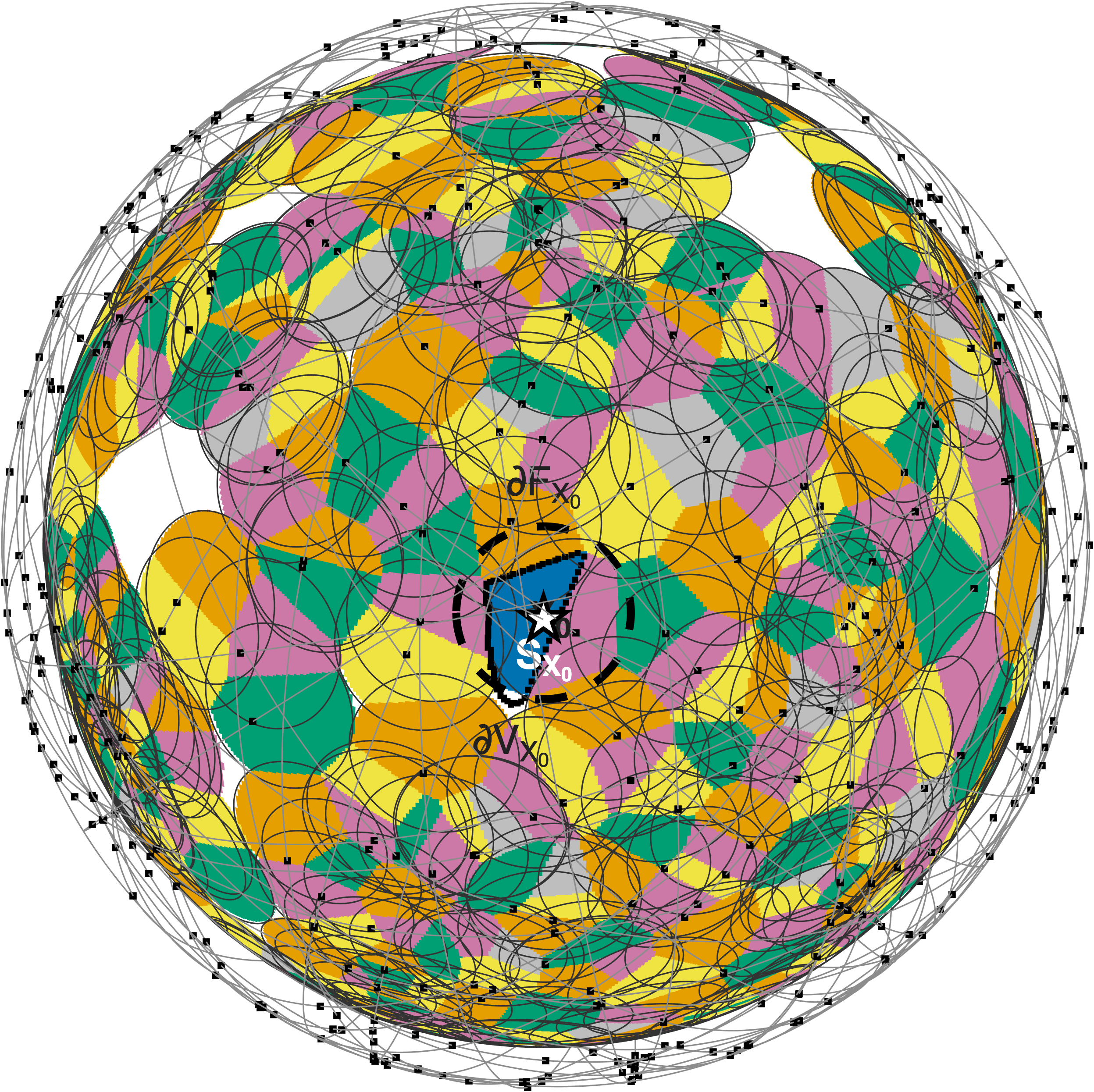}
	\caption{Footprint, spherical Voronoi cell, and service region in one realization of the orbit-aware Cox process ($\gamma_{\rm f}=10^\circ$). Gray great circles are orbits, dots are satellites, thin circles are footprint boundaries, and each colored patch is the service region of one satellite. White patches, where present, are uncovered. For the reference satellite $X_0$ (star), the dashed circle is its footprint boundary $\partial\mathcal F_{X_0}$, the thick curve is its Voronoi cell boundary $\partial\mathcal V_{X_0}(\Phi)$, and the dark patch is its service region $\mathcal S_{X_0}=\mathcal F_{X_0}\cap\mathcal V_{X_0}(\Phi)$. Footprint locations assigned to another satellite take that satellite's color. Voronoi-cell locations outside the footprint are uncovered.}
	\label{fig:system_model}
\end{figure}

Fig.~\ref{fig:system_model} illustrates the three objects in one realization. The footprint $\mathcal F_X$ depends on no other satellite. The Voronoi cell $\mathcal V_X(\Phi)$ is the set of locations for which $X$ is the nearest satellite. The cells form the spherical Voronoi tessellation generated by the satellite directions. Typical-cell statistics are known for uniform point patterns~\cite{sphericalVoronoi,kabluchko2021typical}. A cell may extend beyond the footprint of its generator. Locations of the footprint outside the cell are covered by $X$ but served by a nearer satellite. Locations of the cell outside the footprint are uncovered because every other satellite is farther away and has the same footprint radius. The boundary of a service region therefore consists of footprint-rim portions, where accessibility ends, and of Voronoi edges, where the serving satellite changes. The cells differ from those of a uniform point pattern because same-orbit neighbors and satellites on other orbits have different distributions. Lemma~\ref{lem:palm} in Section~\ref{sec:area_analysis} separates the two groups.

For any measurable $\mathcal A\subseteq\mathbb S_{\rm e}$, let $|\mathcal A|$ denote its spherical surface area and $|\partial\mathcal A|$ the length of its boundary on $\mathbb S_{\rm e}$. The service-region area and perimeter of satellite $X$ are
\begin{equation}
\label{eq:area_perimeter}
	A_X
	=
	|\mathcal S_X(\Phi)|,
	\qquad
	\ell_X
	=
	|\partial\mathcal S_X(\Phi)|,
\end{equation}
respectively.

\subsection{Traffic and Load Model}
\label{subsec:traffic}

We model active users on the Earth surface by a point process $\Upsilon$ on $\mathbb S_{\rm e}$, independent of the satellite process $\Phi$. Unless stated otherwise, $\Upsilon$ is a homogeneous Poisson point process with surface density $\lambda_{\rm u}$ users per unit area. Hence, for any Borel set $B\subseteq\mathbb S_{\rm e}$, the number of active users in $B$ is Poisson with mean $\lambda_{\rm u}|B|$.

For the typical user at $u_0$, the set of accessible satellites consists of those whose footprints contain $u_0$: $	\Phi_{\rm a}(u_0)= \{X\in\Phi:u_0\in\mathcal F_X\}.$
If $\Phi_{\rm a}(u_0)=\emptyset$, the typical user is uncovered. Otherwise, it associates with the nearest accessible satellite,
\begin{equation}
\label{eq:association}
	X^\star
	=
	\argmin_{X\in\Phi_{\rm a}(u_0)}
	\gamma_X(u_0).
\end{equation}
Under this rule, the users assigned to satellite $X$ are precisely those located in $\mathcal S_X(\Phi)$.

The load of satellite $X$ is the number of active users assigned to it:
\begin{equation}
\label{eq:load}
	L_X
	=
	\Upsilon\big(\mathcal S_X(\Phi)\big).
\end{equation}
Conditioned on the satellite geometry,
\begin{equation}
\label{eq:load_poisson}
	L_X\mid\Phi
	\sim
	{\rm Poisson}(\lambda_{\rm u}A_X).
\end{equation}

Thus, the service-region area determines the conditional distribution of satellite load.
 For the typical satellite $X_0$ under the Palm distribution of $\Phi$,
\begin{equation}
\label{eq:mean_load}
	\mathbb E^0[L_{X_0}]
	=
	\lambda_{\rm u}\mathbb E^0[A_{X_0}],
\end{equation}
where $\mathbb E^0[\cdot]$ denotes Palm expectation with respect to the typical satellite.

For user-centric performance, the relevant cell is the service region containing the typical active user, rather than that of the typical satellite. Conditioned on $u_0\in\mathcal B(\Phi)$, let $A^\star=A_{X^\star}$ denote the serving satellite's service-region area. If $N^\star$ is the number of other active users in that region, then
\begin{equation}
\label{eq:typical_user_load}
	N^\star\mid A^\star
	\sim
	{\rm Poisson}(\lambda_{\rm u}A^\star).
\end{equation}
Hence, the serving satellite serves a total of $1+N^\star$ active users, including the typical user.

Alternative association rules can change both the service-region geometry and the load model. Interference-aware selection can choose a different satellite~\cite{kim2025coexistence}. Fixed association biases can produce weighted cells, whereas load-dependent selection can couple the partition to the users. For any rotation-equivariant rule that assigns each covered location to one accessible satellite, the partition identity underlying Corollary~\ref{cor:coverage_load} preserves the typical-satellite mean area. Conservation of the total number of covered users also fixes the population-averaged mean satellite load. However, a user-dependent rule generally invalidates the conditional Poisson load distribution introduced above and the area-bias argument used in the rate analysis. Those calculations require association regions determined independently of the users.

\subsection{Performance Measures}
\label{subsec:performance}

We characterize the transition from the footprint to the association-induced service region through its area and perimeter, and then connect this geometry to the load-aware user rate.

\subsubsection{Service-Region Area}

For the typical satellite $X_0$, the service-region area is $A_{X_0}=|\mathcal S_{X_0}(\Phi)|$. To quantify its size relative to the nominal footprint, we define the normalized service-region area as
\begin{equation}
\label{eq:alpha_def}
	\alpha
	=
	\mathbb E^0\!\left[
	\frac{A_{X_0}}
	{|\mathcal F_{X_0}|}
	\right].
\end{equation}
The quantity $\alpha$ is the mean service-region area relative to the nominal footprint area. Under the traffic model in Section~\ref{subsec:traffic}, the distribution of $A_{X_0}$ determines the corresponding satellite-load distribution.

\subsubsection{Service-Region Perimeter}

For the typical satellite $X_0$, the service-region perimeter is $\ell_{X_0}=|\partial\mathcal S_{X_0}(\Phi)|$. We define its normalized counterpart as
\begin{equation}
\label{eq:beta_def}
	\beta
	=
	\mathbb E^0\!\left[
	\frac{\ell_{X_0}}
	{|\partial\mathcal F_{X_0}|}
	\right].
\end{equation}
The quantity $\beta$ measures mean service-boundary length relative to the nominal footprint boundary. Whereas area determines the traffic assigned to a satellite, perimeter quantifies the total length of the retained footprint boundary and the internal association edges. Converting this descriptor to a handover rate requires additional temporal information.

\subsubsection{Load-Aware User Rate}
Consider a typical user conditioned on being covered by its serving satellite.  Let $C^\star$ denote the
fast-fading-averaged spectral efficiency of the serving link.
The noise-limited propagation model used to evaluate
$C^\star$ will be specified in
Section~\ref{sec:rate_analysis}. The geometric area and
perimeter results are independent of this choice.

We assume that the serving satellite shares its bandwidth $W$ equally among the $1+N^\star$ associated users while maintaining the modeled link SNR. The load-aware user rate is
\begin{equation}
\label{eq:user_rate}
	R^\star
	=
	\frac{W}{1+N^\star}C^\star.
\end{equation}
The mean load-aware rate $\mathbb E[R^\star\mid u_0\in\mathcal B(\Phi)]$ averages over satellite geometry and user load, conditioned on coverage. The unconditional rate, with zero throughput when uncovered, is this quantity multiplied by $P_{\rm cov}^{\rm f}$, the footprint coverage probability defined in Corollary~\ref{cor:coverage_load}. It combines the service-region geometry, which enters through $N^\star$, with the link-level quantity $C^\star$, whose model is specified in Section~\ref{sec:rate_analysis}.
\section{Performance Analysis}
\label{sec:performance_analysis}

\subsection{Analysis of the Normalized Service-Region Area}
\label{sec:area_analysis}

This section characterizes the normalized service-region area $\alpha$ defined in Section~\ref{subsec:performance}. Throughout, all angular distances are Earth-centered. For $v\in\mathbb S_{\rm e}$ and $0<\gamma<\pi/2$, let
\begin{equation}
\label{eq:open_cap}
	\mathcal K(v,\gamma)
	=
	\{x\in\mathbb S_{\rm s}:\gamma_x(v)<\gamma\}
\end{equation}
denote the open spherical cap on the satellite shell containing all satellite positions whose angular separation from $v$ is less than $\gamma$. We assume $0<\gamma_{\rm f}<\pi/2$, as holds under practical altitude and elevation constraints.

For an orbit $\mathcal O$ with unit pole vector
$p_{\mathcal O}$, define its angular offset from
$v\in\mathbb S_{\rm e}$ as
\begin{equation}
	a(\mathcal O,v)
	=
	\inf_{x\in\mathcal O}\gamma_x(v)
	=
	\arcsin\!\left(
	\left|
	p_{\mathcal O}^{\top}\frac{v}{\|v\|}
	\right|
	\right).
	\label{eq:orbit_offset}
\end{equation}
Thus, $\mathcal O$ intersects $\mathcal K(v,\gamma)$ if and
only if $a(\mathcal O,v)<\gamma$.

We begin with the Palm characterization of the orbit-aware Cox process, which specifies the constellation observed from the typical satellite. The characterization was established in~\cite{choi2022analytical,choi2024cox}. We restate it as Lemma~\ref{lem:palm} with a proof outline because the proofs of Theorems~\ref{thm:alpha} and~\ref{thm:beta} use the explicit form of the typical orbit.

\begin{lemma}[Palm distribution of $\Phi$~\cite{choi2022analytical,choi2024cox}]
	\label{lem:palm}
	Let $x_0\in\mathbb S_{\rm s}$, and let $\mathbb P^0$ denote
	the Palm distribution of $\Phi$ with the typical satellite at
	$X_0=x_0$. Under $\mathbb P^0$, the satellite point process
	has the distribution
	\begin{equation}
		\Phi'
		+
		\psi_0
		+
		\delta_{x_0},
		\label{eq:palm_decomposition}
	\end{equation}
	where $\Phi'$ is an independent copy of the original Cox point
	process, and $\psi_0$ is an independent Poisson point process
	of phase intensity $\mu/(2\pi)$ on a random orbit
	$\mathcal O_0\subset\mathbb S_{\rm s}$ passing through $x_0$.
	The unoriented tangent direction of $\mathcal O_0$ at $x_0$ is
	uniformly distributed on $[0,\pi)$.
\end{lemma}

\begin{IEEEproof}
	The process $\Phi$ is a Cox point process directed by the random
	measure
	$M=\sum_i(\mu/2\pi)\nu_i$, where $\nu_i$ is the orbital-phase
	measure on $\mathcal O_i$ satisfying
	$\nu_i(\mathcal O_i)=2\pi$. For a Cox point process, the reduced
	Palm distribution at $x_0$ is obtained from the Palm version of
	its directing random measure~\cite{daley2007introduction}.

	Conditioned on $\Xi$, the satellite processes on distinct orbits
	are independent Poisson point processes. Applying the Mecke
	equation to the Poisson orbit process $\Xi$ and Slivnyak's theorem
	to the satellite process on the selected orbit shows that the
	reduced Palm process consists of an independent copy $\Phi'$ of
	the original Cox process together with an additional Poisson phase
	process $\psi_0$ on an orbit $\mathcal O_0$ passing through $x_0$. See also~\cite{choi2022analytical,choi2024cox}. Adding the Palm
	point $x_0$ gives \eref{eq:palm_decomposition}.

	It remains to characterize $\mathcal O_0$. The isotropic orbit
	measure corresponds to the uniform measure on the space of
	unoriented orbit poles. The orbits through $x_0$ have
	poles satisfying $p^\top x_0=0$, which form a great circle in the
	pole space. Rotation invariance about the axis through $x_0$
	therefore makes the pole of $\mathcal O_0$ uniform on this great
	circle. Equivalently, the unoriented tangent direction of
	$\mathcal O_0$ at $x_0$ is uniform on $[0,\pi)$.
\end{IEEEproof}
Under Lemma~\ref{lem:palm}, the satellites other than $x_0$ form two independent groups, those of the background constellation $\Phi'$ and those on the typical orbit $\mathcal O_0$. The second group is specific to the Palm distribution of the orbit-aware Cox process. Theorem~\ref{thm:alpha} shows that the retention probability of a location in the footprint is the product of the void probabilities of the two groups.

\begin{theorem}[Normalized service-region area]
	\label{thm:alpha}
	For $0\leq a\leq\gamma<\pi/2$, define the orbital-phase
	half-length
	\begin{equation}
		\psi(a,\gamma)
		=
		\arccos\!\left(
		\frac{\cos\gamma}{\cos a}
		\right).
		\label{eq:chordhalf}
	\end{equation}
	Define the inter-orbit retention function as
	\begin{equation}
		\varrho_{\rm inter}(\gamma)
		=
		\exp\!\left(
		-\lambda
		\int_0^{\gamma}
		\left(
		1-e^{-\frac{\mu}{\pi}\psi(a,\gamma)}
		\right)
		\cos a
		\diff a
		\right),
		\label{eq:rho_inter}
	\end{equation}
	and the intra-orbit retention function as
	\begin{equation}
		\varrho_{\rm intra}(\gamma)
		=
		\frac{2}{\pi}
		\int_0^{\pi/2}
		\exp\!\left(
		-\frac{\mu}{\pi}
		\arctan\!\left(
		\tan\gamma\cos\eta
		\right)
		\right)
		\diff\eta.
		\label{eq:rho_intra}
	\end{equation}

	For any $u_\gamma\in\mathcal F_{x_0}$ satisfying
	$\gamma_{x_0}(u_\gamma)=\gamma$, the probability that
	$u_\gamma$ remains assigned to the typical satellite is
	\begin{equation}
		\mathbb P^0
		\big(
		u_\gamma\in\mathcal S_{x_0}(\Phi)
		\big)
		=
		\varrho_{\rm inter}(\gamma)
		\varrho_{\rm intra}(\gamma).
		\label{eq:retention_profile}
	\end{equation}
	Consequently, the mean normalized service-region area of the
	typical satellite is
	\begin{equation}
		\alpha
		=
		\frac{1}{1-\cos\gamma_{\rm f}}
		\int_0^{\gamma_{\rm f}}
		\varrho_{\rm inter}(\gamma)
		\varrho_{\rm intra}(\gamma)
		\sin\gamma
		\diff\gamma.
		\label{eq:thm_alpha}
	\end{equation}
\end{theorem}

\begin{IEEEproof}
	Since the footprint area
	$|\mathcal F_{X_0}|=
	2\pi R_{\rm e}^2(1-\cos\gamma_{\rm f})$
	is deterministic,
	\begin{equation}
		\alpha
		=
		\frac{\mathbb E^0[A_{X_0}]}
		{2\pi R_{\rm e}^2(1-\cos\gamma_{\rm f})}.
		\label{eq:alpha_reduction}
	\end{equation}

	By isotropy, fix $X_0=x_0$. Writing the service-region area
	as an integral over the footprint and applying Fubini's theorem, we obtain
	\begin{equation}
		\begin{aligned}
			\mathbb E^0[A_{X_0}]
			&=
			\mathbb E^0\!\left[
			\int_{\mathcal F_{x_0}}
			\ind_{\{u\in\mathcal V_{x_0}(\Phi)\}}
			\diff u
			\right] \\
			&=
			\int_{\mathcal F_{x_0}}
			\mathbb P^0
			\big(
			u\in\mathcal V_{x_0}(\Phi)
			\big)
			\diff u.
		\end{aligned}
		\label{eq:fubini}
	\end{equation}

	Consider a location $u\in\mathcal F_{x_0}$ with
	$\gamma_{x_0}(u)=\gamma$. The location $u$ belongs to
	$\mathcal V_{x_0}(\Phi)$ if and only if no satellite other than
	$x_0$ lies in $\mathcal K(u,\gamma)$, since this cap contains
	exactly the satellites that are angularly closer to $u$ than
	$x_0$. The use of an open cap excludes $x_0$, which lies on its
	boundary.

	By Lemma~\ref{lem:palm}, the satellites other than $x_0$ form
	the independent superposition $\Phi'+\psi_0$. Hence,
	\begin{equation}
		\begin{aligned}
			\mathbb P^0
			\big(
			u\in\mathcal V_{x_0}(\Phi)
			\big)
			&=
			\underbrace{
				\mathbb P
				\big(
				\Phi'(\mathcal K(u,\gamma))=0
				\big)
			}_{\text{background constellation}} \\
			&\quad\times
			\underbrace{
				\mathbb P
				\big(
				\psi_0(\mathcal K(u,\gamma))=0
				\big)
			}_{\text{typical orbit}}.
		\end{aligned}
		\label{eq:factorization}
	\end{equation}

	Both factors depend on $u$ only through its angular separation
	$\gamma$ from $x_0$. Expressing $\diff u$ in spherical polar
	coordinates centered at the subsatellite point $\bar x_0$ reduces
	\eref{eq:fubini} to
	\begin{equation}
		\begin{aligned}
			\mathbb E^0[A_{X_0}]
			&=
			2\pi R_{\rm e}^2
			\int_0^{\gamma_{\rm f}}
			\mathbb P
			\big(
			\Phi'(\mathcal K_\gamma)=0
			\big) \\
			&\qquad\qquad\times
			\mathbb P
			\big(
			\psi_0(\mathcal K_\gamma)=0
			\big)
			\sin\gamma
			\diff\gamma,
		\end{aligned}
		\label{eq:polar}
	\end{equation}
	where
	$\mathcal K_\gamma=\mathcal K(u_\gamma,\gamma)$
	for an arbitrary $u_\gamma$ satisfying
	$\gamma_{x_0}(u_\gamma)=\gamma$.

	\emph{Inter-orbit void probability.}
	Let $a_i=a(\mathcal O_i,u_\gamma)$ denote the offset of
	$\mathcal O_i$ from the cap center. Under the isotropic orbit
	measure, the pushforward intensity measure of the orbit offsets
	is $\lambda\cos a\,\diff a$ on $[0,\pi/2]$. An orbit intersects
	$\mathcal K_\gamma$ if and only if $a_i<\gamma$.

	For an orbit at offset $a<\gamma$, the spherical Pythagorean
	identity
	$\cos\gamma=\cos a\cos\psi$
	shows that its intersection with $\mathcal K_\gamma$ has phase
	length $2\psi(a,\gamma)$, where $\psi(a,\gamma)$ is given by
	\eref{eq:chordhalf}. Since the satellite phase process has
	intensity $\mu/(2\pi)$, this intersection is empty with
	probability
	$e^{-\frac{\mu}{\pi}\psi(a,\gamma)}$.

	Conditioning on $\Xi$ and applying the probability generating
	functional of the Poisson orbit process, we obtain
	\begin{equation}
		\begin{aligned}
			\mathbb P
			\big(
			\Phi'(\mathcal K_\gamma)=0
			\big)
			&=
			\mathbb E_\Xi\!\left[
			\prod_{i:a_i<\gamma}
			e^{-\frac{\mu}{\pi}\psi(a_i,\gamma)}
			\right] \\
			&=
			\varrho_{\rm inter}(\gamma).
		\end{aligned}
		\label{eq:void_inter}
	\end{equation}

	\emph{Intra-orbit void probability.}
	The typical orbit $\mathcal O_0$ passes through $x_0$, which lies
	on the boundary of $\mathcal K_\gamma$ at angular distance
	$\gamma$ from its center. Let $\eta$ denote the angle at $x_0$
	between the unoriented tangent direction of $\mathcal O_0$ and
	the geodesic from $x_0$ to the cap center. By
	Lemma~\ref{lem:palm}, $\eta$ is uniformly distributed on
	$[0,\pi)$.

	A point of $\mathcal O_0$ at signed phase $\omega$ from $x_0$
	is at angular distance $d(\omega)$ from the cap center, where
	the spherical law of cosines gives
	$\cos d(\omega)
	=
	\cos\omega\cos\gamma
	+
	\sin\omega\sin\gamma\cos\eta$.
	Besides the boundary intersection at $\omega=0$, the second
	solution of $d(\omega)=\gamma$ satisfies
	$\tan(\omega/2)=\tan\gamma\cos\eta$.

	Since $\gamma<\pi/2$, the intersection
	$\mathcal O_0\cap\mathcal K_\gamma$ is a single arc of phase
	length
	$2\arctan(\tan\gamma|\cos\eta|)$. It contains no point of
	$\psi_0$ with probability
	$e^{-\frac{\mu}{\pi}
		\arctan(\tan\gamma|\cos\eta|)}$.
	Averaging over the uniform tangent direction yields
	\begin{equation}
		\begin{aligned}
			\mathbb P
			\big(
			\psi_0(\mathcal K_\gamma)=0
			\big)
			&=
			\frac{2}{\pi}
			\int_0^{\pi/2}
			\exp\!\left(
			-\frac{\mu}{\pi}
			\arctan(
			\tan\gamma\cos\eta
			)
			\right)
			\diff\eta \\
			&=
			\varrho_{\rm intra}(\gamma).
		\end{aligned}
		\label{eq:void_intra}
	\end{equation}
	Equations~\eqref{eq:factorization},
	\eqref{eq:void_inter}, and \eqref{eq:void_intra} establish
	the retention profile in \eref{eq:retention_profile}.
	Substituting the two void probabilities into
	\eref{eq:polar}, and then using \eref{eq:alpha_reduction},
	completes the proof.
\end{IEEEproof}

To evaluate Theorem~\ref{thm:alpha}, we set $\gamma_{\rm f}=10^\circ$ and vary $\lambda\in[10,120]$ and $\mu\in[4,40]$. The analytical surface and the Monte Carlo markers use the same $12\times12$ parameter grid.

\begin{figure}[t]
	\centering
	\includegraphics[width=0.92\linewidth]{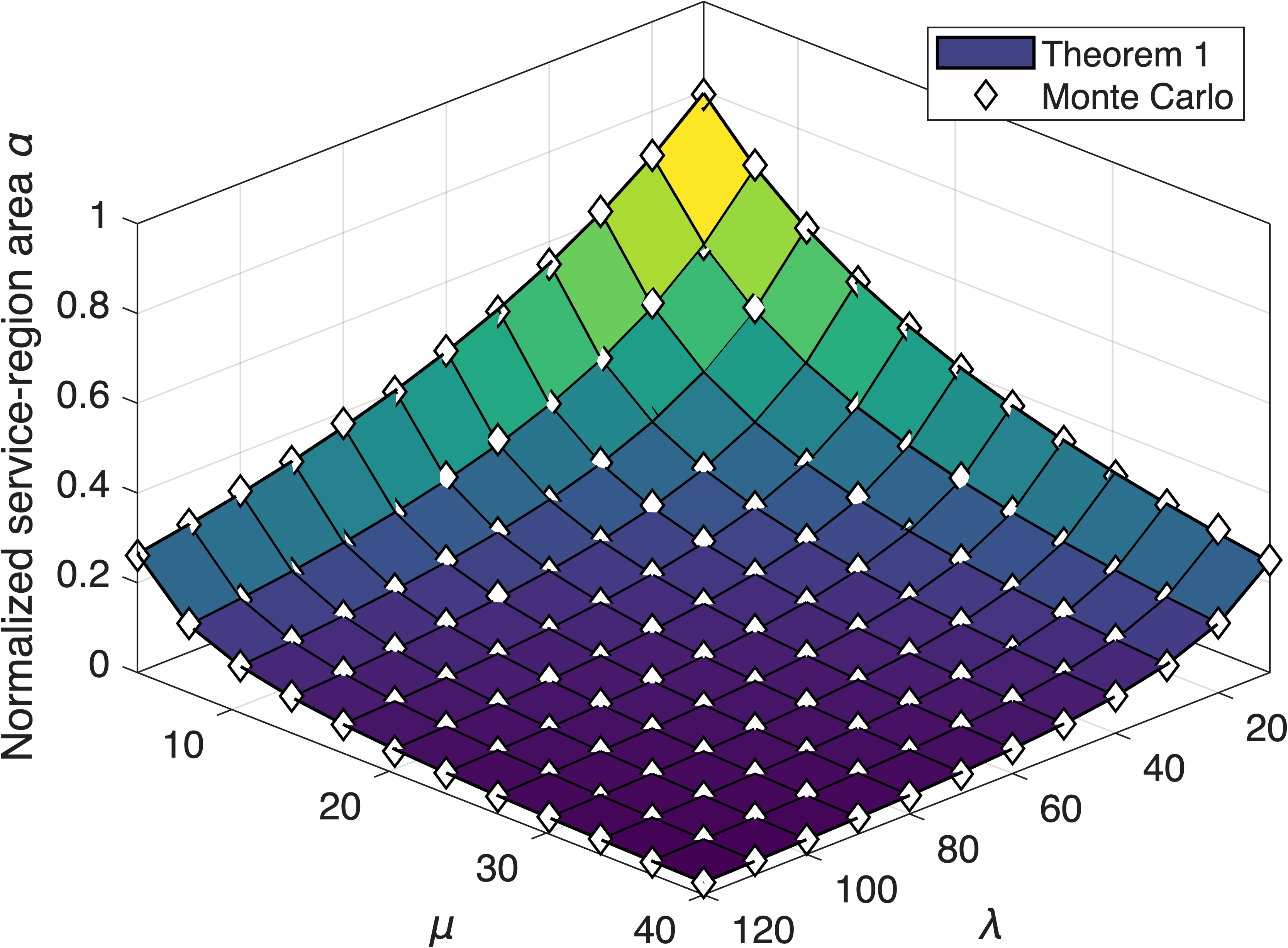}
	\caption{Normalized service-region area versus the mean numbers of orbital planes and satellites per orbit.}
	\label{fig:normalizedserviceregionarea}
\end{figure}

Fig.~\ref{fig:normalizedserviceregionarea} compares Theorem~\ref{thm:alpha} with Monte Carlo simulations. The analytical surface follows the markers over the grid. The value of $\alpha$ decreases with $\lambda$ and with $\mu$: additional orbital planes decrease $\varrho_{\rm inter}$, whereas additional satellites per orbit decrease both $\varrho_{\rm inter}$ and $\varrho_{\rm intra}$. At fixed population, concentrating satellites on fewer orbits reduces coverage and therefore reduces the mean area by Corollary~\ref{cor:coverage_load} below. Indeed, $(1-e^{-\mu x})/\mu$ decreases with $\mu$ for every $x\geq0$, so the coverage exponent decreases when $\lambda\mu$ is fixed. The coverage--load identity therefore determines the net effect of the two retention factors.

\begin{corollary}[From coverage to load]
	\label{cor:coverage_load}
	Let
	$P_{\rm cov}^{\rm f}
	=
	\mathbb P(u_0\in\mathcal B(\Phi))$
	denote the footprint coverage probability of the typical
	user. Then
	$P_{\rm cov}^{\rm f}
	=
	1-\varrho_{\rm inter}(\gamma_{\rm f})$
	and
	\begin{equation}
		\alpha
		=
		\frac{2\,P_{\rm cov}^{\rm f}}
		{\lambda\mu(1-\cos\gamma_{\rm f})}.
		\label{eq:cor_alpha}
	\end{equation}
\end{corollary}

\begin{IEEEproof}
	The typical user is covered if and only if at least one
	satellite lies in $\mathcal K(u_0,\gamma_{\rm f})$. By
	\eref{eq:void_inter} with $\gamma=\gamma_{\rm f}$, the void
	probability of this cap under the unconditioned isotropic process $\Phi$
	is $\varrho_{\rm inter}(\gamma_{\rm f})$, which proves the
	first claim.
	The service regions partition the covered region up to
	boundaries of zero area, so
	$\sum_{X\in\Phi}A_X=|\mathcal B(\Phi)|$.
	By Fubini's theorem and isotropy,
	$\mathbb E[|\mathcal B(\Phi)|]
	=4\pi R_{\rm e}^2P_{\rm cov}^{\rm f}$,
	while the Campbell--Mecke formula gives
	$\mathbb E[\sum_{X\in\Phi}A_X]
	=\lambda\mu\,\mathbb E^0[A_{X_0}]$.
	Hence
	$\mathbb E^0[A_{X_0}]
	=4\pi R_{\rm e}^2P_{\rm cov}^{\rm f}/(\lambda\mu)$,
	and dividing by
	$|\mathcal F_{X_0}|=2\pi R_{\rm e}^2(1-\cos\gamma_{\rm f})$
	yields \eref{eq:cor_alpha}.
\end{IEEEproof}

Corollary~\ref{cor:coverage_load} links the satellite-centric service region to user-centric coverage. It gives the mean satellite load $\mathbb E^0[L_{X_0}]=4\pi R_{\rm e}^2\lambda_{\rm u}P_{\rm cov}^{\rm f}/(\lambda\mu)$ in terms of the footprint coverage probability and the mean population. Once coverage saturates, namely $P_{\rm cov}^{\rm f}\to1$, $\alpha\sim2/(\lambda\mu(1-\cos\gamma_{\rm f}))$ for every allocation of satellites across orbits. Comparing \eqref{eq:cor_alpha} and \eqref{eq:thm_alpha} gives a closed-form identity for the integral of the retention profile. The profile itself gives, in addition, the retention probability at each angular distance $\gamma$, which Theorem~\ref{thm:beta} uses at $\gamma=\gamma_{\rm f}$.

\subsection{Analysis of the Service-Region Perimeter}

We next characterize the normalized service-region perimeter $\beta$. The boundary of $\mathcal S_{X_0}(\Phi)=\mathcal F_{X_0}\cap\mathcal V_{X_0}(\Phi)$ has two geometrically distinct components: the footprint boundary retained by the typical satellite and the Voronoi edges created within the footprint by competing satellites. Theorem~\ref{thm:beta} shows that the first component is determined by the retention profile of Theorem~\ref{thm:alpha} evaluated at the footprint boundary, whereas the second separates into intra-orbit and inter-orbit contributions through the Palm decomposition in Lemma~\ref{lem:palm}. For $x,y\in\mathbb S_{\rm s}$, define their angular separation as $\delta(x,y)=\arccos(x^\top y/R_{\rm s}^2)$ and their spherical bisector on the Earth surface as
\begin{equation}
	\mathcal D(x,y)
	=
	\{u\in\mathbb S_{\rm e}:
	\gamma_x(u)=\gamma_y(u)\}.
	\label{eq:spherical_bisector}
\end{equation}
\begin{theorem}[Normalized service-region perimeter]
	\label{thm:beta}
	For $0\leq a<\pi/2$ and $s\geq0$, define
	\begin{equation}
		\gamma_{\rm b}(a,s)
		=
		\arccos(\cos a\cos s).
		\label{eq:bisector_distance}
	\end{equation}
	Let $\psi(a,\gamma)$, $\varrho_{\rm inter}(\gamma)$,
	and $\varrho_{\rm intra}(\gamma)$ be given in
	Theorem~\ref{thm:alpha}. Then the mean normalized
	service-region perimeter of the typical satellite is
	\begin{equation}
		\beta
		=
		\varrho_{\rm inter}(\gamma_{\rm f})
		\varrho_{\rm intra}(\gamma_{\rm f})
		+
		\frac{T_{\rm intra}+T_{\rm inter}}
		{2\pi\sin\gamma_{\rm f}},
		\label{eq:thm_beta}
	\end{equation}
	where the intra-orbit contribution is
	\begin{equation}
		\begin{aligned}
			T_{\rm intra}
			&=
			\frac{2\mu}{\pi}
			\int_0^{2\gamma_{\rm f}}
			\exp\!\left(
			-\frac{\mu\omega}{2\pi}
			\right) \\
			&\quad\times
			\int_0^{
				\psi(\omega/2,\gamma_{\rm f})
			}
			\varrho_{\rm inter}
			\big(
			\gamma_{\rm b}(\omega/2,s)
			\big)
			\diff s
			\diff\omega,
		\end{aligned}
		\label{eq:T_intra}
	\end{equation}
	and the inter-orbit contribution is
	\begin{equation}
		\begin{aligned}
			T_{\rm inter}
			&=
			\lambda\mu
			\int_0^{2\gamma_{\rm f}}
			\sin\delta
			\int_0^{
				\psi(\delta/2,\gamma_{\rm f})
			} \\
			&\quad\times
			\varrho_{\rm inter}
			\big(
			\gamma_{\rm b}(\delta/2,s)
			\big)
			\varrho_{\rm intra}
			\big(
			\gamma_{\rm b}(\delta/2,s)
			\big)^2
			\diff s
			\diff\delta.
		\end{aligned}
		\label{eq:T_inter}
	\end{equation}
\end{theorem}
\begin{IEEEproof}
	Since the footprint boundary has deterministic length
	$|\partial\mathcal F_{X_0}|
	=
	2\pi R_{\rm e}\sin\gamma_{\rm f}$, we have
	\begin{equation}
		\beta
		=
		\frac{\mathbb E^0[\ell_{X_0}]}
		{2\pi R_{\rm e}\sin\gamma_{\rm f}}.
		\label{eq:beta_reduction}
	\end{equation}
	By isotropy, fix the typical satellite at $X_0=x_0$.
	Up to sets of zero one-dimensional Hausdorff measure, the
	boundary of the service region decomposes as
	\begin{equation}
		\begin{aligned}
			\partial\mathcal S_{x_0}(\Phi)
			&=
			\big(
			\partial\mathcal F_{x_0}
			\cap
			\mathcal V_{x_0}(\Phi)
			\big) \\
			&\quad\cup
			\big(
			\mathcal F_{x_0}
			\cap
			\partial\mathcal V_{x_0}(\Phi)
			\big).
		\end{aligned}
		\label{eq:boundary_decomposition}
	\end{equation}
	The two components intersect in at most finitely many points
	almost surely. Writing $\mathcal H^1$ for the one-dimensional
	Hausdorff measure on $\mathbb S_{\rm e}$, we obtain
	\begin{equation}
		\begin{aligned}
			\mathbb E^0[\ell_{X_0}]
			&=
			\mathbb E^0\!\left[
			\mathcal H^1
			\big(
			\partial\mathcal F_{x_0}
			\cap
			\mathcal V_{x_0}(\Phi)
			\big)
			\right] \\
			&\quad+
			\mathbb E^0\!\left[
			\mathcal H^1
			\big(
			\mathcal F_{x_0}
			\cap
			\partial\mathcal V_{x_0}(\Phi)
			\big)
			\right].
		\end{aligned}
		\label{eq:two_terms}
	\end{equation}

	\emph{Retained footprint boundary.}
	Every $u\in\partial\mathcal F_{x_0}$ satisfies
	$\gamma_{x_0}(u)=\gamma_{\rm f}$. Such a location belongs to
	$\mathcal V_{x_0}(\Phi)$ if and only if no competing satellite
	lies in $\mathcal K(u,\gamma_{\rm f})$. By Fubini's theorem and
	the factorization established in the proof of
	Theorem~\ref{thm:alpha},
	\begin{equation}
		\begin{aligned}
			&\mathbb E^0\!\left[
			\mathcal H^1
			\big(
			\partial\mathcal F_{x_0}
			\cap
			\mathcal V_{x_0}(\Phi)
			\big)
			\right] \\
			&\qquad=
			2\pi R_{\rm e}\sin\gamma_{\rm f}\,
			\varrho_{\rm inter}(\gamma_{\rm f})
			\varrho_{\rm intra}(\gamma_{\rm f}).
		\end{aligned}
		\label{eq:rim_term}
	\end{equation}

	\emph{Voronoi edges inside the footprint.}
	By Lemma~\ref{lem:palm}, under $\mathbb P^0$ the competing
	satellites form the process
	\begin{equation}
		\Phi^{!0}
		\overset{\rm d}{=}
		\Phi'
		+
		\psi_0,
		\label{eq:reduced_palm_competitors}
	\end{equation}
	where $\Phi'$ is an independent copy of the original Cox
	point process and $\psi_0$ is
	the Poisson process on the typical orbit $\mathcal O_0$ through
	$x_0$.
	Almost surely, every nonvertex point
	$u\in
	\mathcal F_{x_0}\cap\partial\mathcal V_{x_0}(\Phi)$
	is equidistant from $x_0$ and exactly one competing satellite
	$Y\in\Phi^{!0}$, with no other satellite strictly closer.
	For $Y\in\Phi^{!0}$, define the exposed bisector segment
	\begin{multline}
			\mathcal E_Y
			=
			\big\{
			u\in
			\mathcal D(x_0,Y)
			\cap
			\mathcal F_{x_0}\\
			\colon
			\big(
			\Phi^{!0}\setminus\{Y\}
			\big)
			\big(
			\mathcal K(
			u,\gamma_{x_0}(u)
			)
			\big)
			=0
			\big\}.
		\label{eq:edge_def}
	\end{multline}
	Up to Voronoi vertices, which form an
	$\mathcal H^1$-null set, the internal service-region boundary
	is the disjoint union of these exposed segments. Therefore,
	\begin{equation}
		\begin{aligned}
			&\mathbb E^0\!\left[
			\mathcal H^1
			\big(
			\mathcal F_{x_0}
			\cap
			\partial\mathcal V_{x_0}(\Phi)
			\big)
			\right] \\
			&\qquad=
			\mathbb E^0\!\left[
			\sum_{Y\in\psi_0}
			\mathcal H^1(\mathcal E_Y)
			\right]
			+
			\mathbb E^0\!\left[
			\sum_{Y\in\Phi'}
			\mathcal H^1(\mathcal E_Y)
			\right].
		\end{aligned}
		\label{eq:edge_split}
	\end{equation}
	We first analyze the geometry common to both terms. Let $Y$
	be a competing satellite at angular separation
	$\delta=\delta(x_0,Y)\in(0,2\gamma_{\rm f})$ from $x_0$.
	Let $m\in\mathbb S_{\rm e}$ be the Earth location in the
	direction of the midpoint of the shorter geodesic from
	$\widehat x_0=x_0/R_{\rm s}$ to $\widehat Y=Y/R_{\rm s}$. It is the point of $\mathcal D(x_0,Y)$ closest to the
	subsatellite point $\bar x_0$.
	For a point $u\in\mathcal D(x_0,Y)$ at signed angular
	displacement $s$ from $m$, the geodesic from $\bar x_0$
	to $m$ meets the bisector orthogonally. The spherical
	Pythagorean identity therefore gives
\begin{equation}
		\gamma_{x_0}(u)
		=
		\gamma_{\rm b}(\delta/2,s)
		=
		\arccos\!\left(
		\cos(\delta/2)\cos s
		\right).
		\label{eq:bisector_param}
	\end{equation}
	The condition $u\in\mathcal F_{x_0}$ is equivalent to
	\begin{equation}
		|s|
		\leq
		\psi(\delta/2,\gamma_{\rm f}).
		\label{eq:bisector_footprint_range}
	\end{equation}
	Thus, the bisector does not intersect the footprint unless
	$\delta<2\gamma_{\rm f}$. Moreover, for every point on this
	bisector segment, both $x_0$ and $Y$ lie on the boundary of
	the competitor cap
	$\mathcal K(u,\gamma_{x_0}(u))$.

	\emph{Intra-orbit contribution.}
	Condition on the direction of the typical orbit $\mathcal O_0$,
	and let $Y\in\psi_0$ have absolute phase separation
	$\omega\in(0,\pi)$ from $x_0$. Since both satellites lie on
	the same great circle, $\delta(x_0,Y)=\omega$. The bisector
	intersects the footprint only when
	$\omega<2\gamma_{\rm f}$.
	For a point $u$ on the corresponding bisector segment, the
	intersection of $\mathcal O_0$ with
	$\mathcal K(u,\gamma_{x_0}(u))$ is the open minor arc between
	$x_0$ and $Y$, whose phase length is $\omega$. Applying the
	Mecke equation to the Poisson phase process $\psi_0$, under
	which the remaining points of $\psi_0$ form a Poisson phase process
	of the same intensity, gives
	\begin{equation}
		\begin{aligned}
			\mathbb E^0\!\left[
			\sum_{Y\in\psi_0}
			\mathcal H^1(\mathcal E_Y)
			\right]
			&=
			\frac{\mu}{2\pi}
			\cdot2
			\int_0^{2\gamma_{\rm f}}
			\exp\!\left(
			-\frac{\mu\omega}{2\pi}
			\right)
			\cdot2R_{\rm e} \\
			&\hphantom{={}}\times
			\int_0^{
				\psi(\omega/2,\gamma_{\rm f})
			}
			\varrho_{\rm inter}
			\big(
			\gamma_{\rm b}(\omega/2,s)
			\big)
			\diff s
			\diff\omega \\
			&=
			R_{\rm e}T_{\rm intra}.
		\end{aligned}
		\label{eq:intra_term}
	\end{equation}
	Here, the first factor of two accounts for the two phase
	directions from $x_0$, and the second accounts for the two
	sides of the bisector segment. The exponential term is the
	void probability of the remaining typical-orbit satellites on
	the minor arc between $x_0$ and $Y$, while
	$\varrho_{\rm inter}$ is the void probability of the
	independent process $\Phi'$. The integrand does not depend on
	the direction of $\mathcal O_0$, so the conditioning can be
	dropped.

	\emph{Inter-orbit contribution.}
	The first-order intensity of $\Phi'$ on the unit direction
	sphere is $\lambda\mu/(4\pi)$ with respect to its surface
	measure $\sigma$. We apply the Campbell--Mecke formula to
	$\Phi'$ at a candidate satellite $y$: by
	Lemma~\ref{lem:palm} applied at $y$, under the reduced Palm
	distribution of $\Phi'$ at $y$ the process is again
	distributed as $\Phi'+\psi_y$, where $\psi_y$ is a Poisson
	phase process on a uniformly directed orbit through $y$,
	and $\Phi'$, $\psi_y$, and $\psi_0$ are mutually
	independent. Consequently,
	\begin{equation}
		\begin{aligned}
			&\mathbb E^0\!\left[
			\sum_{Y\in\Phi'}
			\mathcal H^1(\mathcal E_Y)
			\right]
			=
			\frac{\lambda\mu}{4\pi}
			\int_{\mathbb S^2}
			\mathbb E\!\left[
			\mathcal H^1(\mathcal E_y)
			\right]
			\sigma(\diff\widehat y),
		\end{aligned}
		\label{eq:campbell_cox}
	\end{equation}
	where the void condition in $\mathcal E_y$ is evaluated for
	the process $\Phi'+\psi_y+\psi_0$.
	For a point $u$ on the bisector between $x_0$ and $y$, let
	$\gamma=\gamma_{x_0}(u)=\gamma_y(u)$.
The process
	$\Phi'$ avoids $\mathcal K(u,\gamma)$ with probability
	$\varrho_{\rm inter}(\gamma)$. The satellite processes on the orbits through $x_0$
	and $y$ independently avoid this
	cap with probability $\varrho_{\rm intra}(\gamma)$ each,
	since $x_0$ and $y$ both lie on the boundary of
	$\mathcal K(u,\gamma)$ and each orbit has a uniform direction
	at its base point, exactly as in the proof of
	Theorem~\ref{thm:alpha}. The square in \eref{eq:T_inter}
	arises from these two independent orbit
	contributions.
	By isotropy,
	$\sigma(\diff\widehat y)=2\pi\sin\delta\,\diff\delta$.
	Using \eref{eq:bisector_param} and the arc-length element
	$R_{\rm e}\diff s$, we obtain
\begin{equation}
	\begin{aligned}
		\mathbb E^0\!\left[
		\sum_{Y\in\Phi'}
		\mathcal H^1(\mathcal E_Y)
		\right]
		&=
		\frac{\lambda\mu}{4\pi}\cdot 2\pi
		\int_0^{2\gamma_{\rm f}}
		\sin\delta\cdot 2R_{\rm e} \\
		&\hphantom{={}}\times
		\int_0^{\psi(\delta/2,\gamma_{\rm f})}
		\varrho_{\rm inter}\big(\gamma_{\rm b}(\delta/2,s)\big) \\
		&\hphantom{={}}\times
		\varrho_{\rm intra}\big(\gamma_{\rm b}(\delta/2,s)\big)^2
		\,\diff s\,\diff\delta \\
		&=
		R_{\rm e}\,T_{\rm inter}.
	\end{aligned}
	\label{eq:inter_term}
\end{equation}
	Finally, we combine \eref{eq:two_terms}, \eref{eq:rim_term}, \eref{eq:intra_term}, and \eref{eq:inter_term}. Dividing the result by $2\pi R_{\rm e}\sin\gamma_{\rm f}$, as specified in \eref{eq:beta_reduction}, yields \eref{eq:thm_beta}.
\end{IEEEproof}

\emph{Numerical evaluation.} We tabulate the intra-orbit retention function and the inter-orbit exponent on an angular grid using one-dimensional quadrature. These tables depend on $\mu$ and the footprint radius. We obtain $\varrho_{\rm inter}$ for each $\lambda$ by exponentiation. We then evaluate \eref{eq:T_intra} and \eref{eq:T_inter} using two-dimensional quadrature with interpolation of the tabulated functions. For $N_\gamma$ angular points, $N_a$ offset points, and $N_\eta$ direction points, tabulation costs $O(N_\gamma(N_a+N_\eta))$ operations. Each perimeter contribution requires $O(N_{\rm outer}N_{\rm inner})$ interpolations. We check angular-grid refinement and the independent area identity \eref{eq:cor_alpha}.
 Grid refinement changes the normalized perimeter by less than $0.03\%$ in the tested configurations. The area integral agrees with \eref{eq:cor_alpha} to within $0.004\%$.

\begin{remark}[Composition of the service boundary]
	\label{rem:beta_composition}
	The first term in \eref{eq:thm_beta} is the retention profile of Theorem~\ref{thm:alpha} at the rim, and the second collects the Voronoi edges created within the footprint, separated by Lemma~\ref{lem:palm} according to their generating satellites. In $T_{\rm intra}$, the exponential factor is the void probability of the minor arc between two same-orbit satellites, so the dominant same-orbit edges arise from neighbors at the along-orbit spacing $2\pi/\mu$. In $T_{\rm inter}$, the factor $\varrho_{\rm intra}$ is squared because the orbits through the two satellites must both avoid the exposed cap. Neither feature appears in Poisson or binomial spherical models. The service region is a geodesically convex subset of a cap smaller than a hemisphere. The spherical Crofton formula expresses its perimeter through intersections with great circles~\cite{chiu2013stochastic}. Such a great circle intersects a convex boundary twice or not at all, except for a null set of tangencies. Containment therefore bounds its perimeter by that of the footprint, giving $\beta\leq1$.
\end{remark}

\begin{remark}[Perimeter scaling and association boundaries]
\label{rem:beta_scaling}
The area satisfies $\alpha\sim2/[\lambda\mu(1-\cos\gamma_{\rm f})]$ when coverage tends to one. The perimeter does not have a universal scaling in $\lambda\mu$. The numerical slopes approach $-1/2$ for increasing $\lambda$ at fixed $\mu$ and for the balanced path $\lambda/\mu=1.5$. This observation does not establish that exponent for every path along which both parameters grow.

At fixed $\lambda$, dominated convergence gives $\varrho_{\rm inter}(\gamma)\to e^{-\lambda\sin\gamma}$ as $\mu\to\infty$. In $T_{\rm intra}$, the substitution $t=\mu\omega/(2\pi)$ concentrates the bisector separation at zero and yields $T_{\rm intra}\to4\int_0^{\gamma_{\rm f}}e^{-\lambda\sin s}\,\diff s$. The retained rim and the inter-orbit contribution vanish. Thus,
\begin{equation}
\lim_{\mu\to\infty}\beta=
\frac{2}{\pi\sin\gamma_{\rm f}}\int_0^{\gamma_{\rm f}}e^{-\lambda\sin s}\,\diff s
\approx\frac{2}{\pi\lambda\sin\gamma_{\rm f}},
\label{eq:beta_floor}
\end{equation}
where the approximation requires large $\lambda$ and $\lambda\sin\gamma_{\rm f}\gg1$. Same-orbit neighbors constrain the along-orbit width of the cell. Its long sides are their bisectors, which extend across the typical orbit. The mean extent in that direction remains limited by the other orbital planes and the footprint.

The sum of cell perimeters counts each internal association edge twice and each retained footprint rim once. Consequently, the mean length of distinct internal edges per unit Earth area is
\begin{equation}
\mathcal D_{\rm V}=\frac{\lambda\mu(T_{\rm intra}+T_{\rm inter})}{8\pi R_{\rm e}}.
\label{eq:boundary_density}
\end{equation}
This spatial quantity excludes coverage boundaries. A temporal crossing rate additionally requires the normal velocity of each edge relative to the user. Satellite ground-track speed alone does not specify that velocity, so we do not infer a handover rate from the perimeter.
\end{remark}

\begin{figure}[t]
	\centering
	\includegraphics[width=0.92\linewidth]{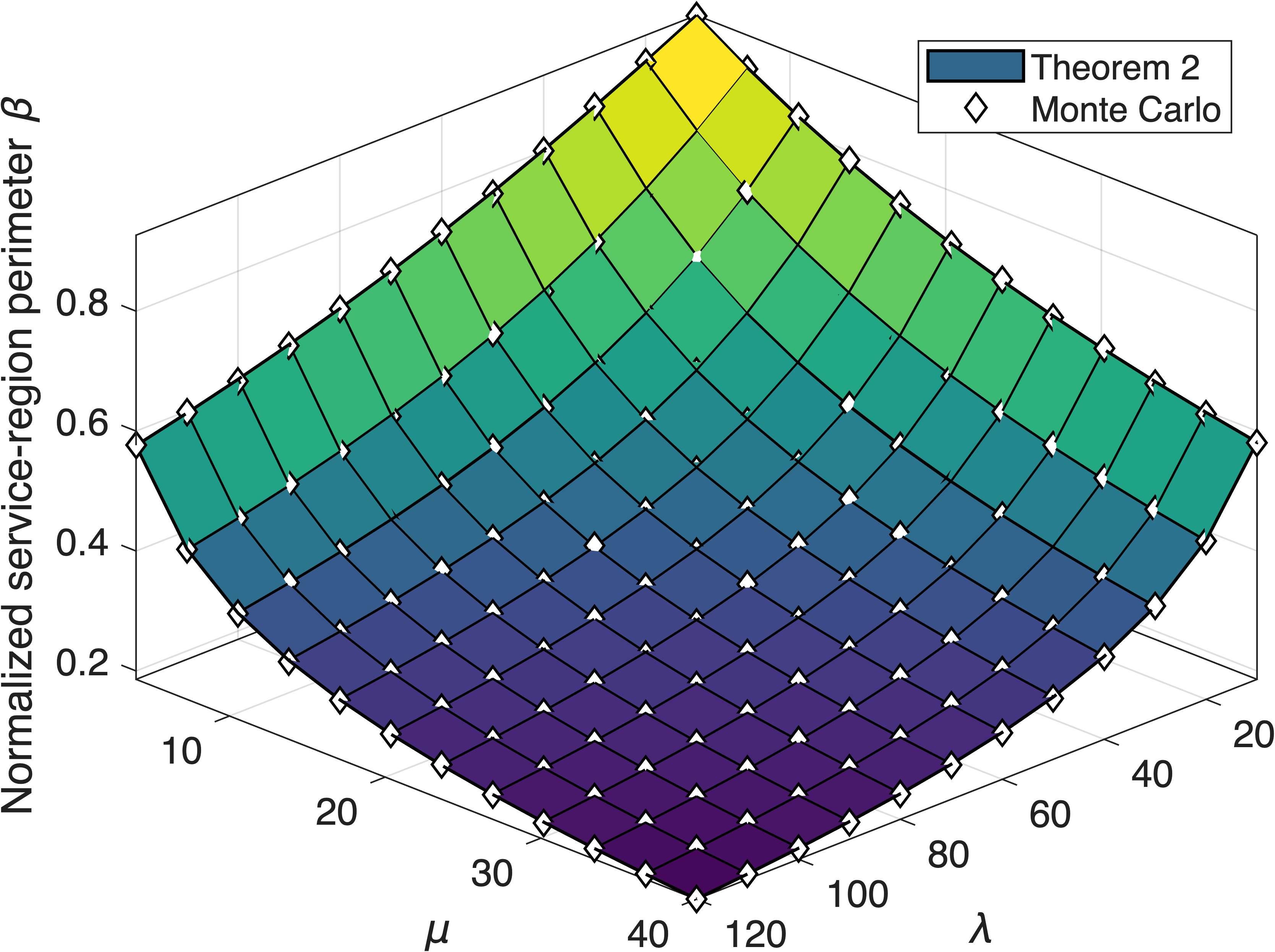}
	\caption{Normalized service-region perimeter: analysis and Monte Carlo validation.}
	\label{fig:perimeter_surface}
\end{figure}

\begin{figure}[t]
	\centering
	\includegraphics[width=0.92\linewidth]{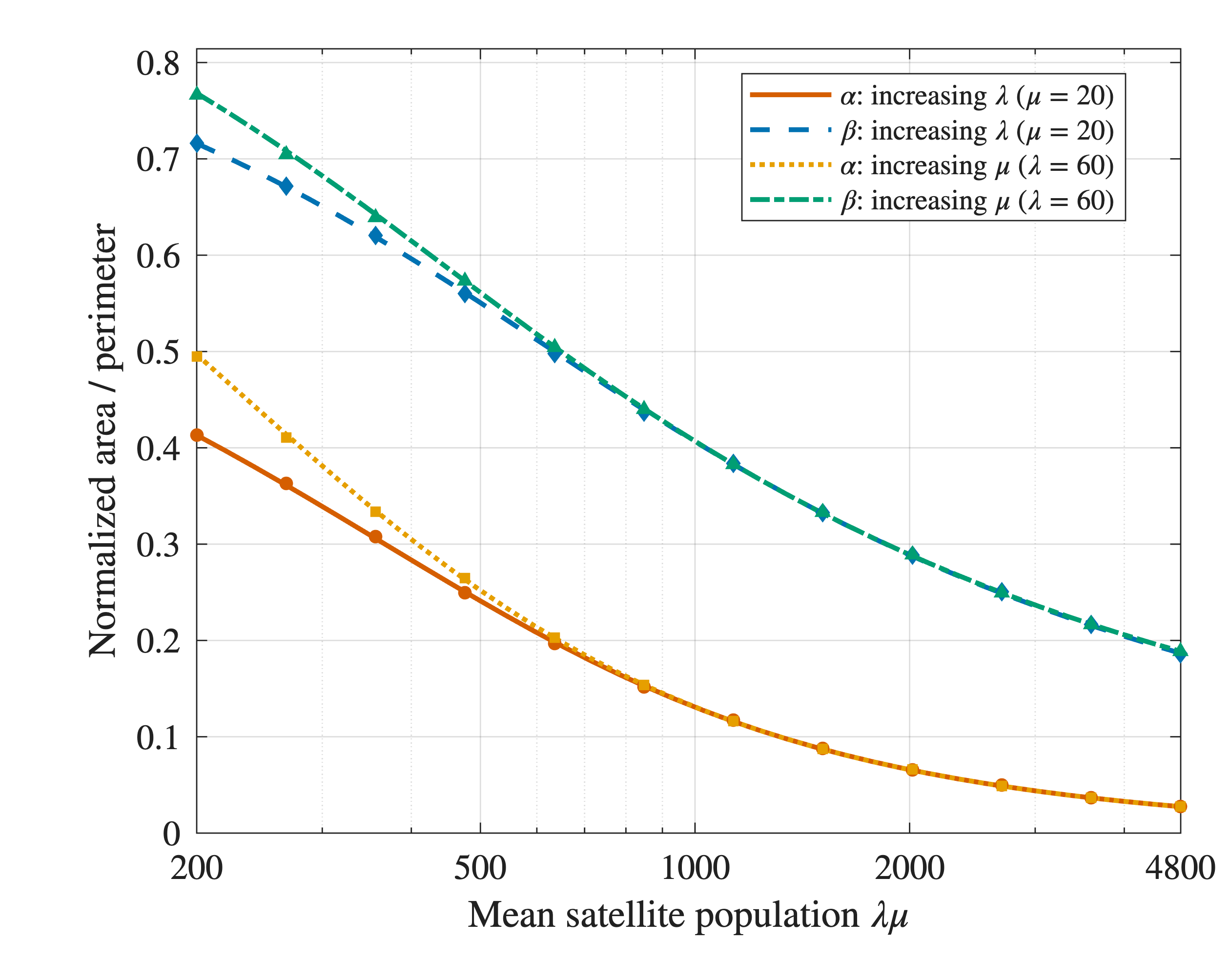}
	\caption{Normalized area and perimeter under constellation densification. Curves show Theorems~\ref{thm:alpha} and~\ref{thm:beta}. Markers show Monte Carlo estimates from independent Palm-cell realizations.}
	\label{fig:area_perimeter_densification}
\end{figure}

\begin{figure}[t]
	\centering
	\includegraphics[width=0.92\linewidth]{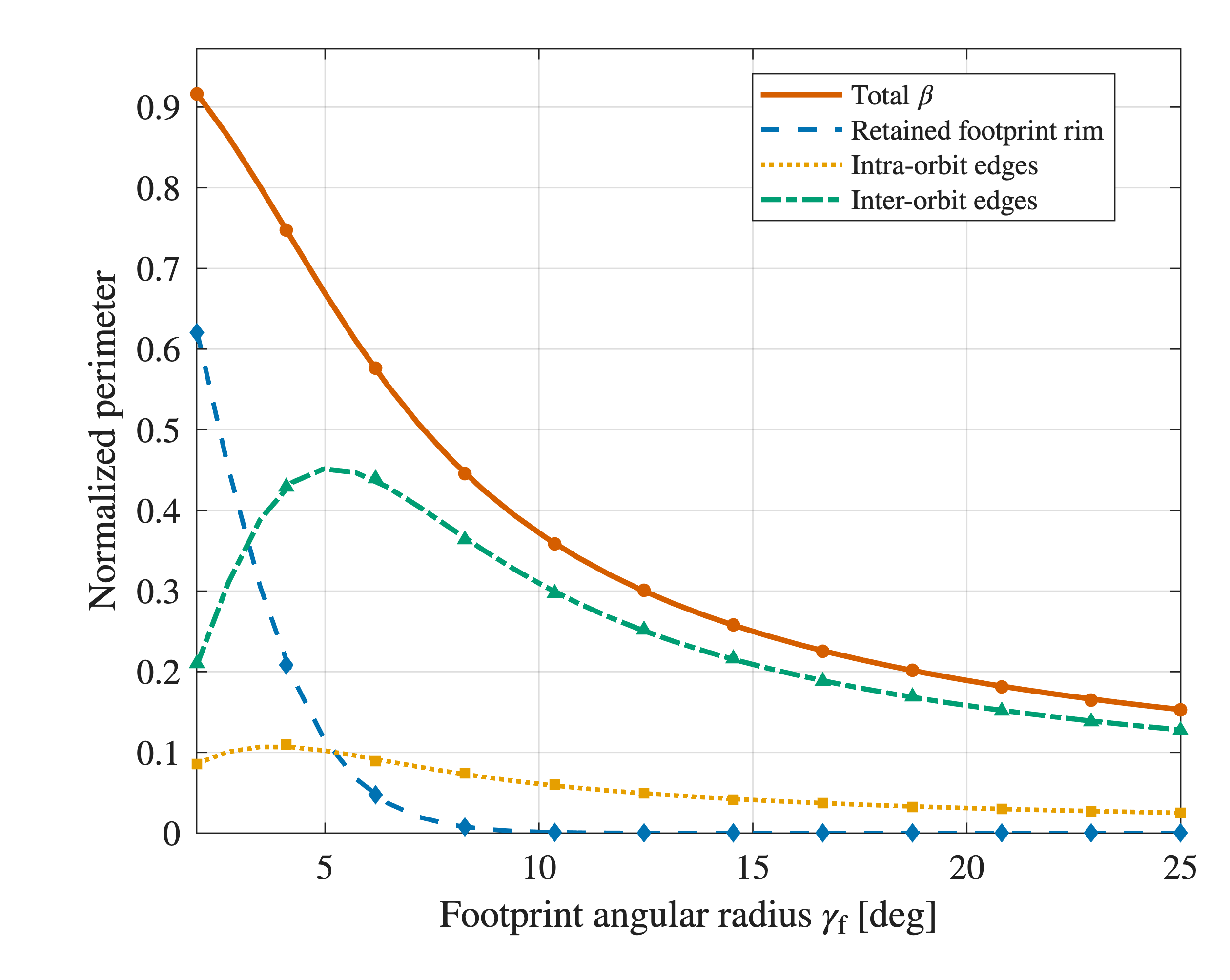}
	\caption{Normalized service-boundary components versus footprint radius. Curves show Theorem~\ref{thm:beta}. Markers show Monte Carlo estimates of the total perimeter and its three components.}
	\label{fig:footprint_radius_perimeter_decomposition}
\end{figure}

We evaluate Theorem~\ref{thm:beta} in Fig.~\ref{fig:perimeter_surface} using $\gamma_{\rm f}=10^\circ$ and the same $\lambda$ and $\mu$ grid as in Fig.~\ref{fig:normalizedserviceregionarea}. In Fig.~\ref{fig:area_perimeter_densification}, we keep $\gamma_{\rm f}=10^\circ$ and vary the mean population from $200$ to $4{,}800$ along two paths, $\mu=20$ with increasing $\lambda$ and $\lambda=60$ with increasing $\mu$. In Fig.~\ref{fig:footprint_radius_perimeter_decomposition}, we fix $(\lambda,\mu)=(60,20)$ and vary $\gamma_{\rm f}$ from $2^\circ$ to $25^\circ$. This last plot is a geometric sensitivity sweep. At $h=550$~km, the zero-elevation horizon is about $23^\circ$, so its largest radii are not feasible footprints at that altitude.

Fig.~\ref{fig:perimeter_surface} compares Theorem~\ref{thm:beta} with Monte Carlo simulations. The analytical surface follows the markers over the grid. The normalized perimeter decreases with $\lambda$ and with $\mu$. Adding satellites contracts the existing satellite's convex service region, which cannot increase its perimeter.

Fig.~\ref{fig:area_perimeter_densification} compares the normalized area and perimeter as the mean population increases along the two paths. Along both plotted paths, the mean area and hence the mean satellite load decrease faster than the mean perimeter. The difference between the two paths is largest in the sparse-to-intermediate regime. In the higher-population part of the plotted range, both the area and perimeter curves are close across the two paths. The area curves nearly coincide as both coverage probabilities approach one, consistent with Corollary~\ref{cor:coverage_load}. The perimeter nevertheless has different asymptotic behavior along the two paths. When $\mu$ increases at $\lambda=60$, its positive limit in \eref{eq:beta_floor} is not yet reached within the plotted range.

Fig.~\ref{fig:footprint_radius_perimeter_decomposition} separates normalized perimeter into retained-footprint, intra-orbit, and inter-orbit components. For a small footprint, a large part of the service boundary coincides with the footprint rim. As the footprint radius increases, the rim contribution vanishes and the boundary consists of Voronoi edges, among which the inter-orbit component dominates while the intra-orbit component remains smaller but not negligible. Enlarging the footprint therefore does not enlarge the service boundary proportionally: beyond the radius at which the rim contribution vanishes, the boundary is determined by association alone.

\subsection{Analysis of the Load-Aware User Rate}
\label{sec:rate_analysis}

We now characterize the mean load-aware rate of the typical covered user defined in Section~\ref{subsec:performance}. Two random quantities determine the rate: the serving-link spectral efficiency $C^\star$, governed by the serving distance, and the number $N^\star$ of other active users sharing the serving satellite, governed by the service-region area. Lemma~\ref{lem:serving_distance} gives the distribution of the serving distance, and Theorem~\ref{thm:rate} combines it with the service-region area of Theorem~\ref{thm:alpha}.

\begin{lemma}[Serving-distance density conditioned on coverage~\cite{choi2022analytical}]
	\label{lem:serving_distance}
	Let $\gamma^\star=\gamma_{X^\star}(u_0)$ denote the angular
	separation between the typical covered user and its serving
	satellite. Conditioned on $u_0\in\mathcal B(\Phi)$, the
	density of $\gamma^\star$ is
	\begin{equation}
		f_{\gamma^\star}(\gamma)
		=
		\frac{\lambda\mu}{2P_{\rm cov}^{\rm f}}\,
		\varrho_{\rm inter}(\gamma)
		\varrho_{\rm intra}(\gamma)
		\sin\gamma,
		\quad
		0\leq\gamma\leq\gamma_{\rm f}.
		\label{eq:serving_density}
	\end{equation}
\end{lemma}

\begin{IEEEproof}
	The typical user is covered if and only if its nearest
	satellite lies within angle $\gamma_{\rm f}$, in which case
	that satellite is the serving satellite. For a measurable
	function $g\geq0$,
	\begin{equation}
		\begin{aligned}
			&\mathbb E\!\left[
			g(\gamma^\star)\,
			\ind_{\{u_0\in\mathcal B(\Phi)\}}
			\right] 
			\;=
			\mathbb E\bigg[
			\sum_{X\in\Phi}
			g\big(\gamma_X(u_0)\big)\,
			\ind_{\{\gamma_X(u_0)\leq\gamma_{\rm f}\}} \\
			&\hspace{4.5em}\times
			\ind_{\{(\Phi\setminus\{X\})(\mathcal K(u_0,\gamma_X(u_0)))=0\}}
			\bigg].
		\end{aligned}
		\label{eq:serving_campbell_setup}
	\end{equation}
	The mean measure of $\Phi$ assigns mass
	$(\lambda\mu/2)\sin\gamma\diff\gamma$ to the set of satellite
	positions at angular separation $\gamma$ from $u_0$. By the
	Campbell--Mecke formula and Lemma~\ref{lem:palm} applied at a
	position $x_\gamma$ with
	$\gamma_{x_\gamma}(u_0)=\gamma$, the reduced Palm process is
	distributed as $\Phi'+\psi_{x_\gamma}$, where $x_\gamma$ lies
	on the boundary of the cap
	$\mathcal K(u_0,\gamma)$---the same configuration as in the
	proof of Theorem~\ref{thm:alpha}. The corresponding void
	probability therefore factorizes into
	$\varrho_{\rm inter}(\gamma)\varrho_{\rm intra}(\gamma)$, and we have
	\begin{multline}
		\mathbb E\!\left[
		g(\gamma^\star)
		\ind_{\{u_0\in\mathcal B(\Phi)\}}
		\right]\\
		=\frac{\lambda\mu}{2}
		\int_0^{\gamma_{\rm f}}
		g(\gamma)
		\varrho_{\rm inter}(\gamma)
		\varrho_{\rm intra}(\gamma)
		\sin\gamma
		\diff\gamma.
		\label{eq:serving_campbell}
	\end{multline}
	Setting $g\equiv1$ recovers
	$P_{\rm cov}^{\rm f}$, consistent with
	Corollary~\ref{cor:coverage_load}. Dividing
	\eref{eq:serving_campbell} by $P_{\rm cov}^{\rm f}$ yields
	\eref{eq:serving_density}.
\end{IEEEproof}

The serving distance is the contact distance of~\cite{choi2022analytical} conditioned on footprint coverage. Let $\gamma_{\rm c}$ denote the angular separation to the nearest satellite. Then $\gamma^\star$ has the distribution of $\gamma_{\rm c}$ given $\gamma_{\rm c}\leq\gamma_{\rm f}$, so \eref{eq:serving_density} is the truncated and normalized contact-distance density. The identity
$-\mathrm d\varrho_{\rm inter}/\mathrm d\gamma=(\lambda\mu/2)\varrho_{\rm inter}\varrho_{\rm intra}\sin\gamma$
gives the same representation. Thus the product $\varrho_{\rm inter}\varrho_{\rm intra}$ determines both the mean service-region area in \eref{eq:thm_alpha} and the serving-distance density. We use this connection to express the load-aware rate through the same retention functions.

For the serving link, we adopt a normalized, noise-limited propagation model with distance-dependent attenuation and fading. The slant range at angular separation $\gamma$ is
\begin{equation}
	d(\gamma)
	=
	\sqrt{R_{\rm e}^2+R_{\rm s}^2-2R_{\rm e}R_{\rm s}\cos\gamma},
	\label{eq:slant_range}
\end{equation}
with $d(0)=h$. All deterministic link-budget terms, including transmit power, antenna gains, carrier-dependent free-space loss, and receiver noise power, are absorbed into $\mathrm{SNR}_0$, the mean received signal-to-noise ratio at nadir. Before small-scale fading, the mean SNR at angular separation $\gamma$ is modeled as $\mathrm{SNR}_0(h/d(\gamma))^\kappa$, where $\kappa$ is the effective path-loss exponent. Within the footprint, the effective antenna gain is constant and absorbed into $\mathrm{SNR}_0$; outside the footprint, the satellite is inaccessible. 
A common location-dependent gain $G(\gamma)$ with $G(0)=1$ enters \eref{eq:link_map} through $\mathrm{SNR}_0G(\gamma)(h/d(\gamma))^\kappa$. At a fixed footprint radius, this changes the spectral efficiency but not the service-region geometry or load distribution. We keep the prescribed footprint fixed in the gain sensitivity calculation below and vary the radial gain profile within it.

Let $H$ denote the fading power gain of the serving link, normalized to $\mathbb E[H]=1$ and independent of $\Phi$ and $\Upsilon$. For LEO downlinks, $H$ is commonly modeled as shadowed-Rician~\cite{ppp_extensions}. The fast-fading-averaged spectral efficiency of the serving link, for the constant-gain reference, is
\begin{align}
	&C^\star
	=
	C(\gamma^\star),
	\\
	&C(\gamma)
	=
	\mathbb E_H\!\left[
	\log_2\!\left(
	1+
	\mathrm{SNR}_0
	\left(
	\frac{h}{d(\gamma)}
	\right)^{\!\kappa}
	H
	\right)
	\right].
	\label{eq:link_map}
\end{align}
The fading average in \eref{eq:link_map} admits the single-integral
representation
\begin{equation}
	C(\gamma)
	=
	\frac{1}{\ln 2}
	\int_0^\infty
	\frac{e^{-z}}{z}
	\Big(
	1-
	\mathcal L_H\big(
	z\,\mathrm{SNR}_0
	(h/d(\gamma))^{\kappa}
	\big)
	\Big)
	\diff z,
	\label{eq:frullani}
\end{equation}
where $\mathcal L_H(s)=\mathbb E[e^{-sH}]$ is the Laplace
transform of $H$, which is available in closed form for the
shadowed-Rician model. The no-fading baseline is recovered by
setting $H\equiv1$.
The load $N^\star$ is driven by the service-region area $A^\star$ of the serving satellite. A closed-form area distribution is unavailable even for planar Poisson--Voronoi cells, whose areas are accurately approximated by Gamma distributions~\cite{gammaApprox}.

\begin{assumption}
	\label{ass:rate}
	(i)~We approximate the typical service-region area by a Gamma distribution with shape $k>0$ and mean $\mathbb E^0[A_{X_0}]$ given by Corollary~\ref{cor:coverage_load}, where $k$ is a fitted model parameter. (ii)~Conditioned on $u_0\in\mathcal B(\Phi)$, the fast-fading-averaged spectral efficiency $C(\gamma^\star)$ of the serving link and the service-region area $A^\star$ are independent.
\end{assumption}

Theorems~\ref{thm:alpha} and~\ref{thm:beta}, Corollary~\ref{cor:coverage_load}, and Lemma~\ref{lem:serving_distance} are exact for the model of Section~\ref{sec:system_model}. Theorem~\ref{thm:rate} holds under Assumption~\ref{ass:rate} and is therefore an approximation for the model itself. We examine the two parts of the assumption separately after the theorem.
The Gamma distribution cannot be exact because the service-region area satisfies $0<A_{X_0}\leq|\mathcal F_{X_0}|$, whereas a Gamma distribution has unbounded support. A service region that coincides with the entire footprint also contributes an atom at the upper endpoint. We use it as a load approximation and assess its accuracy separately from the independence assumption.

\begin{theorem}[Load-aware user rate under Assumption~\ref{ass:rate}]
	\label{thm:rate}
	Let
	$\bar L
	=
	\lambda_{\rm u}\mathbb E^0[A_{X_0}]
	=
	4\pi R_{\rm e}^2\lambda_{\rm u}
	P_{\rm cov}^{\rm f}/(\lambda\mu)$
	denote the mean load of the typical satellite. Under
	Assumption~\ref{ass:rate}, the mean rate of the typical
	covered user is
	\begin{equation}
		\mathbb E[R^\star\mid u_0\in\mathcal B(\Phi)]
		=
		W\bar C\,
		\frac{1-\left(1+\bar L/k\right)^{-k}}{\bar L},
		\label{eq:thm_rate}
	\end{equation}
	where
	\begin{equation}
		\bar C
		=
		\frac{\lambda\mu}{2P_{\rm cov}^{\rm f}}
		\int_0^{\gamma_{\rm f}}
		C(\gamma)\,
		\varrho_{\rm inter}(\gamma)
		\varrho_{\rm intra}(\gamma)
		\sin\gamma
		\diff\gamma.
		\label{eq:mean_C}
	\end{equation}
\end{theorem}

\begin{IEEEproof}
	Conditioned on coverage, $R^\star=WC(\gamma^\star)/(1+N^\star)$, where $C$ already averages over the fading gain.
	Because the user process is independent of the satellite
	geometry, $N^\star$ and $C(\gamma^\star)$ are conditionally
	independent given $A^\star$. Together with
	Assumption~\ref{ass:rate}(ii), this implies the factorization below.
	\begin{equation}
		\mathbb E[R^\star\,|\,u_0\in\mathcal B(\Phi)]
		=
		W\,
		\mathbb E[C(\gamma^\star)]\,
		\mathbb E\!\left[\frac{1}{1+N^\star}\right],
		\label{eq:rate_decoupled}
	\end{equation}
	where both expectations are conditioned on coverage. The first
	factor equals $\bar C$ by
	Lemma~\ref{lem:serving_distance}.

	For the second factor, since
	$N^\star\,|\,A^\star\sim
	{\rm Poisson}(\lambda_{\rm u}A^\star)$,
	\begin{equation}
		\mathbb E\!\left[
		\frac{1}{1+N^\star}
		\,\Big|\,
		A^\star
		\right]
		=
		\sum_{n\geq0}
		\frac{e^{-\lambda_{\rm u}A^\star}
			(\lambda_{\rm u}A^\star)^n}{(n+1)!}
		=
		\frac{1-e^{-\lambda_{\rm u}A^\star}}
		{\lambda_{\rm u}A^\star}.
		\label{eq:poisson_inverse}
	\end{equation}
	The service region containing the typical covered user is the
	area-biased version of the typical service region. This size
	bias arises because the typical covered user samples service
	regions by surface area rather than uniformly by satellite. Consequently, a service region of area $a$ is selected with
	probability proportional to $a$. Under the Gamma approximation in Assumption~\ref{ass:rate}(i), its area density
	is
	\[
	f_{A^\star}(a) = \frac{a f_A(a)}{\mathbb E^0[A_{X_0}]},
	\]
	where $f_A$ is the density of the typical service-region area.
	Under Assumption~\ref{ass:rate}(i), $f_A$ is the
	${\rm Gamma}(k,r)$ density with rate parameter
	$r=k/\mathbb E^0[A_{X_0}]$, so the area-biased density $f_{A^\star}(a)$ is
	${\rm Gamma}(k+1,r)$. Consequently,
	\begin{equation}
		\begin{aligned}
			&\mathbb E\!\left[
			\frac{1-e^{-\lambda_{\rm u}A^\star}}
			{\lambda_{\rm u}A^\star}
			\right]\\
			&=
			\frac{r^{k+1}}{\lambda_{\rm u}\Gamma(k+1)}
			\int_0^\infty
			a^{k-1}
			\big(
			1-e^{-\lambda_{\rm u}a}
			\big)
			e^{-ra}
			\diff a \\
			&=
			\frac{r}{\lambda_{\rm u}k}
			\left[
			1-
			\left(
			\frac{r}{r+\lambda_{\rm u}}
			\right)^{\!k}
			\right] =
			\frac{1-\left(1+\bar L/k\right)^{-k}}{\bar L},
		\end{aligned}
		\label{eq:load_factor}
	\end{equation}
	where the last step uses
	$\lambda_{\rm u}/r=\bar L/k$. Substituting $\bar C$ and
	\eref{eq:load_factor} into \eref{eq:rate_decoupled} yields
	\eref{eq:thm_rate}.
\end{IEEEproof}

The noise-limited model isolates geometric resource sharing. For the same serving satellite, resource allocation, desired signal, and noise, removing nonnegative interference gives a pointwise upper bound on the instantaneous rate. Its exact ensemble mean therefore bounds the corresponding interference-aware mean. Theorem~\ref{thm:rate} approximates that noise-limited mean and is not itself a guaranteed upper bound, because the Gamma approximation and link--load factorization can err in either direction.

The load factor in \eref{eq:thm_rate} decreases with $\bar L$ and increases with $k$. As $k\to\infty$, it approaches $(1-e^{-\bar L})/\bar L$, the value obtained when every service region has the deterministic area $\mathbb E^0[A_{X_0}]$. As $\bar L\to\infty$, it behaves as $1/\bar L$, yielding $\mathbb E[R^\star \mid u_0\in\mathcal B(\Phi)]\approx W\bar C/\bar L$. The conditional mean rate is therefore approximately inversely proportional to mean satellite load. We estimate the shape parameter from Palm samples of the typical service-region area. Below, we assess the resulting Gamma approximation separately from the independence approximation in Assumption~\ref{ass:rate}. The value $k=2$ is an empirical reference for the balanced allocations examined here.

For the rate experiments, we set
$R_{\rm e}=6371$~km, $h=550$~km,
$\gamma_{\rm f}=10^\circ$, $\kappa=2$,
$\mathrm{SNR}_0=10$~dB, $W=100$~MHz,
$\lambda_{\rm u}=10^{-5}$~users/km$^2$, and $k=2$.
The serving link follows the average-shadowing shadowed-Rician model with $(b,m,\Omega)=(0.126,10.1,0.835)$, where $2b$, $m$, and $\Omega$ denote average scattered power, shadowing severity, and average line-of-sight power, respectively~\cite{abdi2003new}. The fading power gain is normalized so that $\mathbb E[H]=1$. For the parameter-grid experiments, we use $\lambda\in[20,200]$ and $\mu\in[10,100]$. The wider density range relative to Figs.~\ref{fig:normalizedserviceregionarea}--\ref{fig:footprint_radius_perimeter_decomposition} exposes the light-load saturation predicted by Theorem~\ref{thm:rate}. To compare orbital allocations at the same total population, we also consider three densification paths for $c\in\{2^{0},2^{1/2},\ldots,2^{7/2}\}$: balanced, $(\lambda,\mu)=(15c,10c)$; plane-heavy, $(15c^2,10)$; and orbit-heavy, $(15,10c^2)$. All three have the same population $\lambda\mu=150c^2$.

We assess the two parts of Assumption~\ref{ass:rate} separately using Monte Carlo simulations on a $6\times6$ grid with $\lambda\in[20,200]$ and $\mu\in[10,100]$ at $\gamma_{\rm f}=10^\circ$. We also use eight balanced-path configurations at each of $8^\circ$, $10^\circ$, and $15^\circ$. We obtain Palm areas by radial boundary integration. We estimate the service-region area $A^\star$ from uniform footprint samples and then evaluate the exact conditional Poisson resource share. This joint user-rate estimator does not impose either part of Assumption~\ref{ass:rate}, but finite area sampling introduces numerical error.

\begin{figure}[t]
\centering
\includegraphics[width=0.92\linewidth]{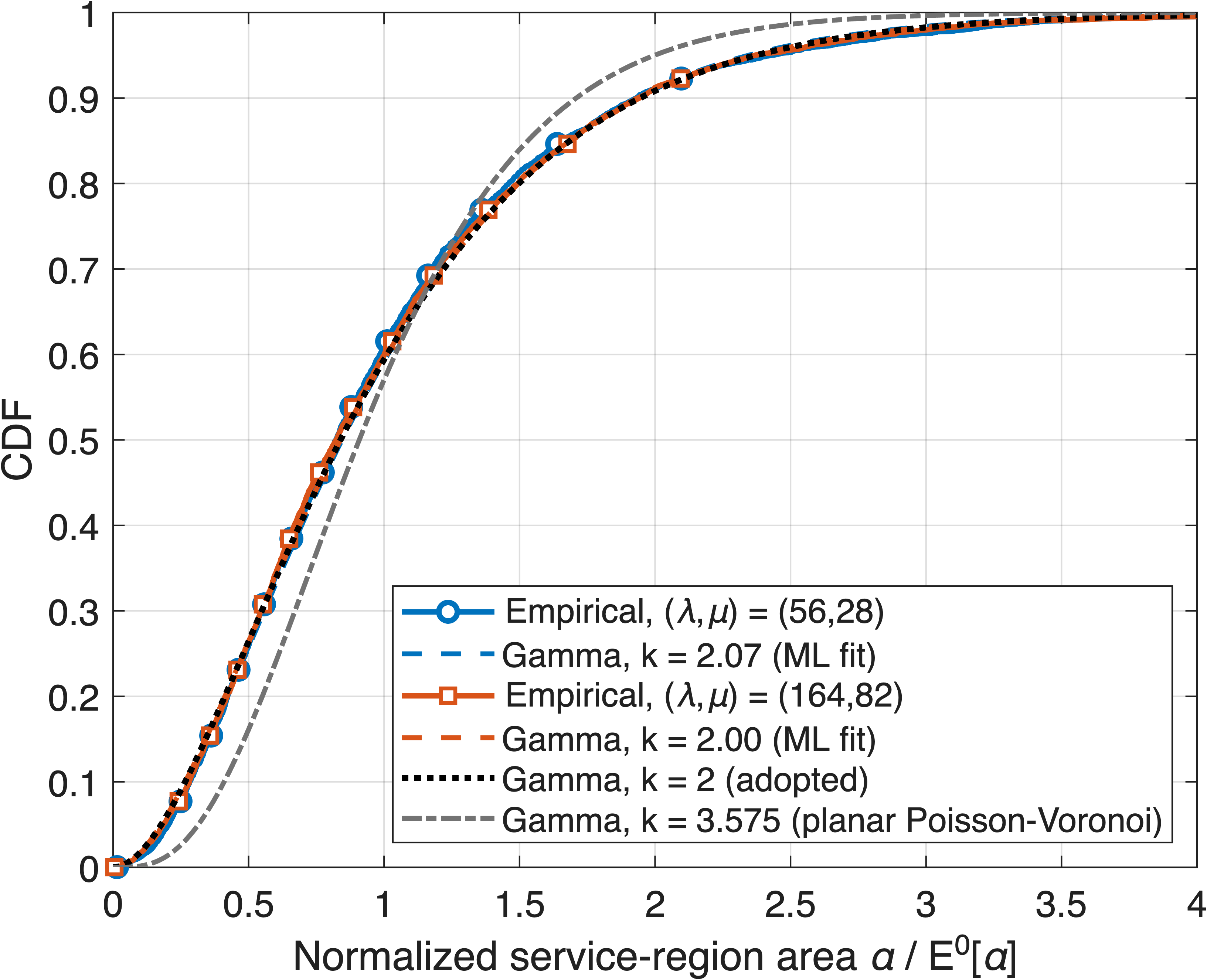}
\caption{Empirical distribution of $A_{X_0}/\mathbb E^0[A_{X_0}]$ for $(\lambda,\mu)=(56,28)$ and $(164,82)$, compared with fitted Gamma distributions and the reference shapes $k=2$ and $k=3.575$ ($\gamma_{\rm f}=10^\circ$).}
\label{fig:gamma_area_cdf}
\end{figure}
The fitted shapes are 2.07 and 2.00 in Fig.~\ref{fig:gamma_area_cdf}. In the coverage-saturated part of the grid they range from 1.24 to 2.93, so $k=2$ is a useful balanced-allocation approximation rather than a universal constant. The value 3.49 at $(20,10)$ reflects a different, partially footprint-constrained regime. The planar reference $k\approx3.575$ is itself a fitted approximation~\cite{pineda2004cell}. Many sparsely populated orbits reduce orbital dependence, but we do not assert an exact Gamma limiting distribution.

We assess the Gamma contribution by comparing its load factor with the area-biased empirical load factor from Palm samples. With $k=2$, the largest relative discrepancy is 0.087 over the same configurations.

We next assess the independence approximation through its effect on the mean rate.
For $\ell(a)=(1-e^{-\lambda_{\rm u}a})/(\lambda_{\rm u}a)$, define the dependence error by
\begin{equation}
\varepsilon_{\rm ind}=\frac{\mathbb E[C(\gamma^\star)\ell(A^\star)]}{\mathbb E[C(\gamma^\star)]\mathbb E[\ell(A^\star)]}-1,
\label{eq:dependence_error}
\end{equation}
with all expectations conditioned on coverage. Across the 60 configurations, the Pearson correlation between spectral efficiency and service-region area $A^\star$ ranges from approximately $-0.50$ to $-0.20$. Thus the variables are not independent. The measured $|\varepsilon_{\rm ind}|$ is at most 0.088. At $(164,82)$ it is 0.002, although the correlation is $-0.42$. There, the conditional mean spectral efficiency lies between 3.04 and 3.13 bit/s/Hz across the area quintiles. Concentration of the serving distance limits the rate effect of dependence in that dense configuration.

The maximum total rate discrepancy against the joint Monte Carlo estimator is 0.086. These are finite-sample comparisons, not analytical error bounds. In particular, the two discrepancies use different empirical samples and do not add to the total rate error.

\begin{figure}[t]
\centering
\includegraphics[width=0.92\linewidth]{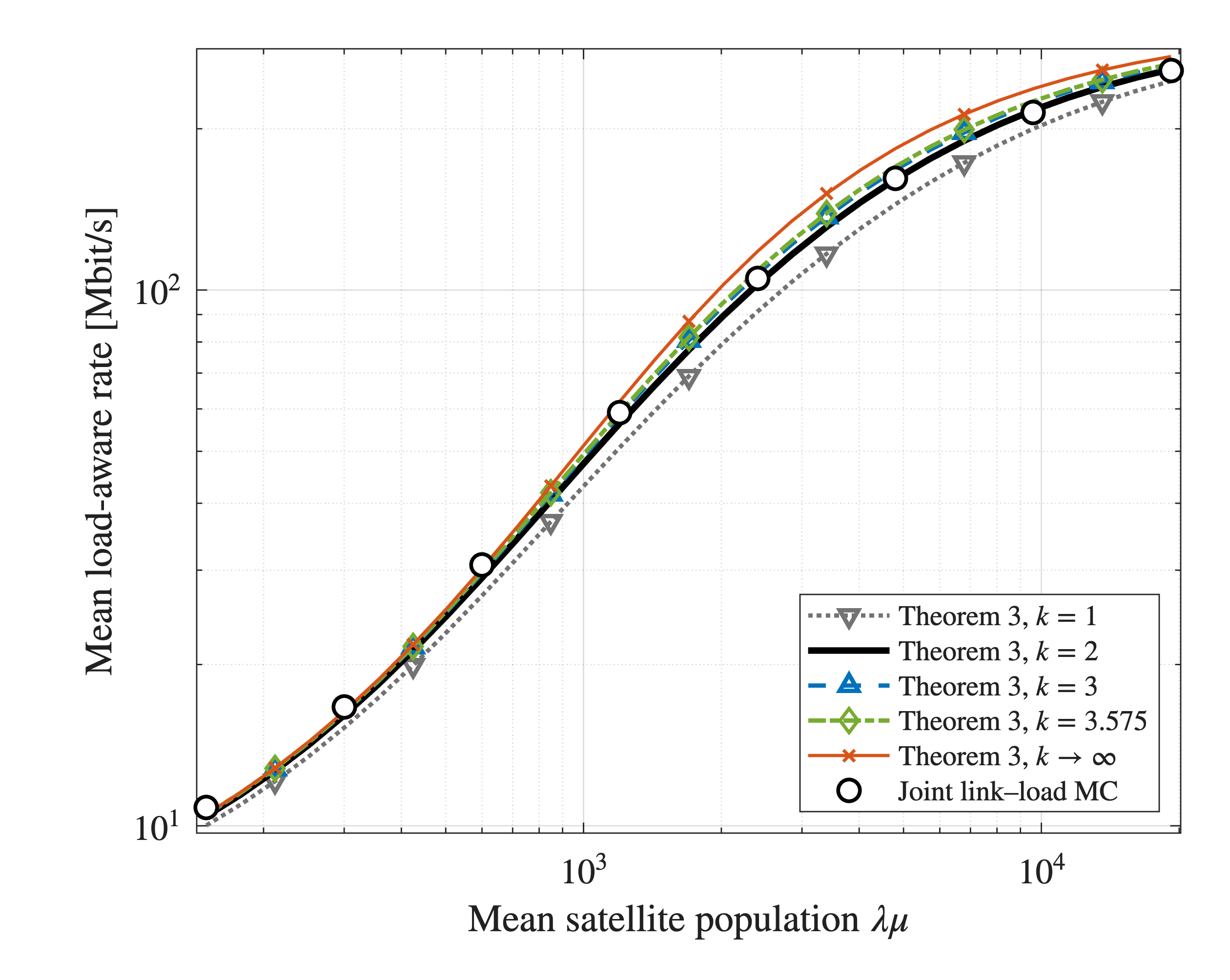}
\caption{Sensitivity of the approximate rate to $k\in\{1,2,3,3.575\}$ and $k\to\infty$ along the balanced densification path, with a joint link--load Monte Carlo estimate ($\gamma_{\rm f}=10^\circ$).}
\label{fig:rate_k_sensitivity}
\end{figure}
Fig.~\ref{fig:rate_k_sensitivity} shows that area dispersion matters most at intermediate load. With $k=2$, the analytical values differ from the Monte Carlo estimates by at most 6\% along this path and by less than 3\% from population 2,400 onward. The deterministic-area curve can exceed those estimates by about 14\%. The curves for $k=1$ and $k\to\infty$ delimit the displayed Gamma family with $k\geq1$; they are not bounds on arbitrary service-region distributions or dependent link--load models.

We also compare footprint sampling with radial boundary integration on the same constellation draws for eight Cox and Walker cases. The mean rate difference is below $0.7\%$ in every case, and angular refinement has a negligible effect. The Walker discrepancies discussed below therefore persist when we replace the footprint-sampling estimator.

\begin{figure}[t]
\centering
\includegraphics[width=0.92\linewidth]{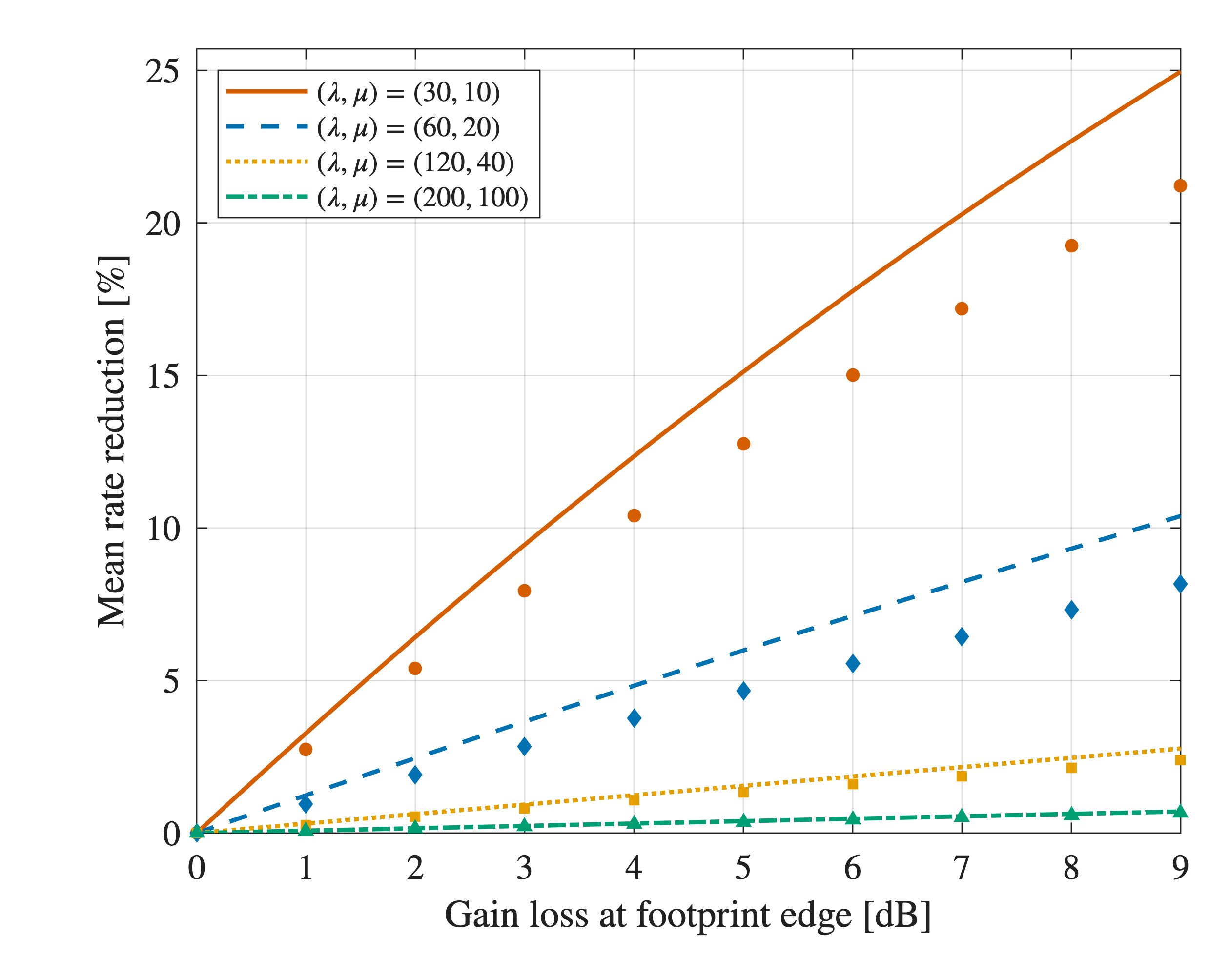}
\caption{Mean-rate reduction under the common radial gain $G(\gamma)=10^{-b_{\rm e}(\gamma/\gamma_{\rm f})^2/10}$, where $b_{\rm e}$ is the edge loss in dB. We hold the footprint radius at $10^\circ$. Curves show Theorem~\ref{thm:rate} with the gain-dependent link factor. Markers show joint link--load Monte Carlo estimates.}
\label{fig:beam_sensitivity}
\end{figure}
Fig.~\ref{fig:beam_sensitivity} evaluates gain roll-off without changing accessibility. At 3-dB edge loss, the reductions in the link factor, and therefore in the approximate rate, are 9.45\%, 3.65\%, and 0.93\% for $(\lambda,\mu)=(30,10),(60,20),(120,40)$. The reduction becomes smaller as the angular serving distances decrease. The joint link--load Monte Carlo estimates give reductions of 7.95\%, 2.84\%, and 0.81\%, respectively. The difference reflects the link--load factorization in the approximate rate. The service-region area, perimeter, and load distribution remain unchanged in this comparison.

\begin{figure}[t]
\centering
\includegraphics[width=0.92\linewidth]{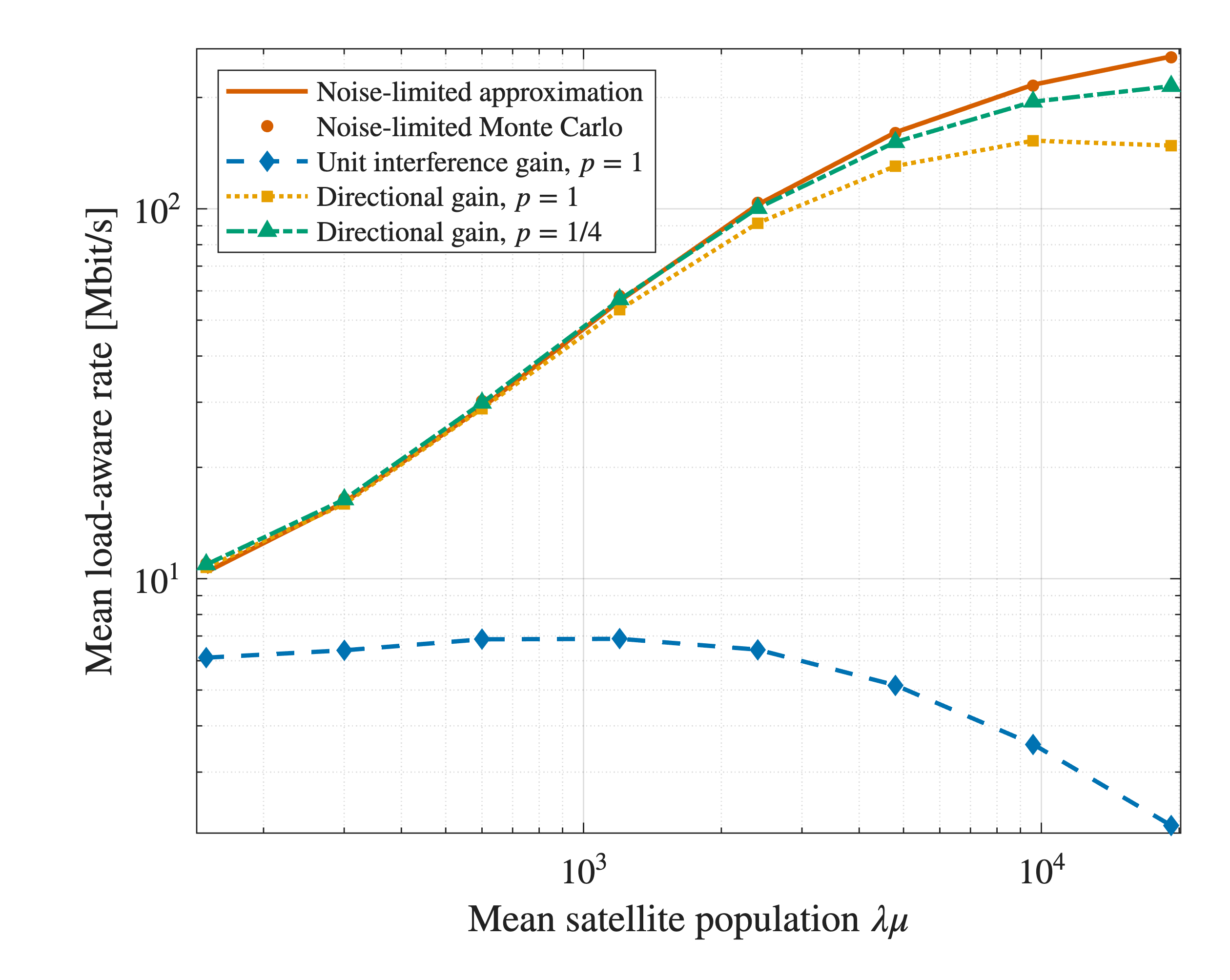}
\caption{Load-aware rate along the balanced path with independent interferer activity $p=1$ or $p=1/4$ and a directional or unit-gain interference pattern. The serving bandwidth and desired-link SNR are fixed ($\gamma_{\rm f}=10^\circ$).}
\label{fig:sinr_rate_densification}
\end{figure}

\begin{remark}[Load factor under densification]
	\label{rem:densification}
	Theorem~\ref{thm:rate} factorizes user rate into a link factor $\bar C$ and a load factor. As the constellation densifies, the serving-distance distribution in \eref{eq:serving_density} concentrates near zero, but $\bar C$ remains bounded by the nadir spectral efficiency $C(0)$. By contrast, after coverage saturates and while $\bar L\gg1$, the load factor behaves as $1/\bar L\propto\lambda\mu$. Along the balanced path from $(\lambda,\mu)=(15,10)$ to $(120,80)$, $\bar C$ increases by a factor of $1.43$, whereas the load factor increases by a factor of $14.5$. The rate gain under densification therefore comes from the decrease of the mean load. At population $9{,}600$, the mean satellite load is approximately $0.53$, and the load factor is still below one. Further densification has diminishing returns. The limiting noise-limited value is $WC(0)$ along paths on which the serving distance tends to zero and load vanishes. The mechanism presumes that the aggregate co-channel interference at the user terminal remains below the noise. Fig.~\ref{fig:sinr_rate_densification} quantifies the interference penalty at the tested densities.
\end{remark}

\begin{remark}[Footprint radius after coverage saturation]
	\label{rem:footprint}
	In a dense constellation, the service region is association-limited and the footprint boundary is rarely retained ($\varrho_{\rm inter}(\gamma_{\rm f})\varrho_{\rm intra}(\gamma_{\rm f})\approx0$ in Theorem~\ref{thm:beta}). Enlarging the footprint therefore provides negligible additional gain in the dense regime: at $(\lambda,\mu)=(72,22)$, the parameters of the Starlink-type shell of Section~\ref{subsec:walker}, increasing $\gamma_{\rm f}$ from $8^\circ$ to $20^\circ$ changes the mean rate by less than $1\%$. 
Conversely, reducing $\gamma_{\rm f}$ to $6^\circ$ raises the mean rate conditioned on coverage by about $5\%$. The serving-distance distribution shifts toward shorter links and the mean load decreases, but coverage falls to $96\%$.

	For the noise-limited equal-share model, a useful footprint design reference is the smallest $\gamma_{\rm f}$ for which the probability that a user is uncovered satisfies $\varrho_{\rm inter}(\gamma_{\rm f})\leq\epsilon$. Enlarging the accessible footprint beyond this point provides little additional rate benefit because association already limits the service region.

\end{remark}

The active-user density $\lambda_{\rm u}$ is an exogenous traffic parameter that enters Theorem~\ref{thm:rate} only through the mean load $\bar L=\lambda_{\rm u}\mathbb E^0[A_{X_0}]$. For the Monte Carlo comparisons, we reuse the same realizations of $(\gamma^\star,A^\star)$ and re-evaluate the load factor at each user density.

\begin{figure}[t]
	\centering
	\includegraphics[width=0.92\linewidth]{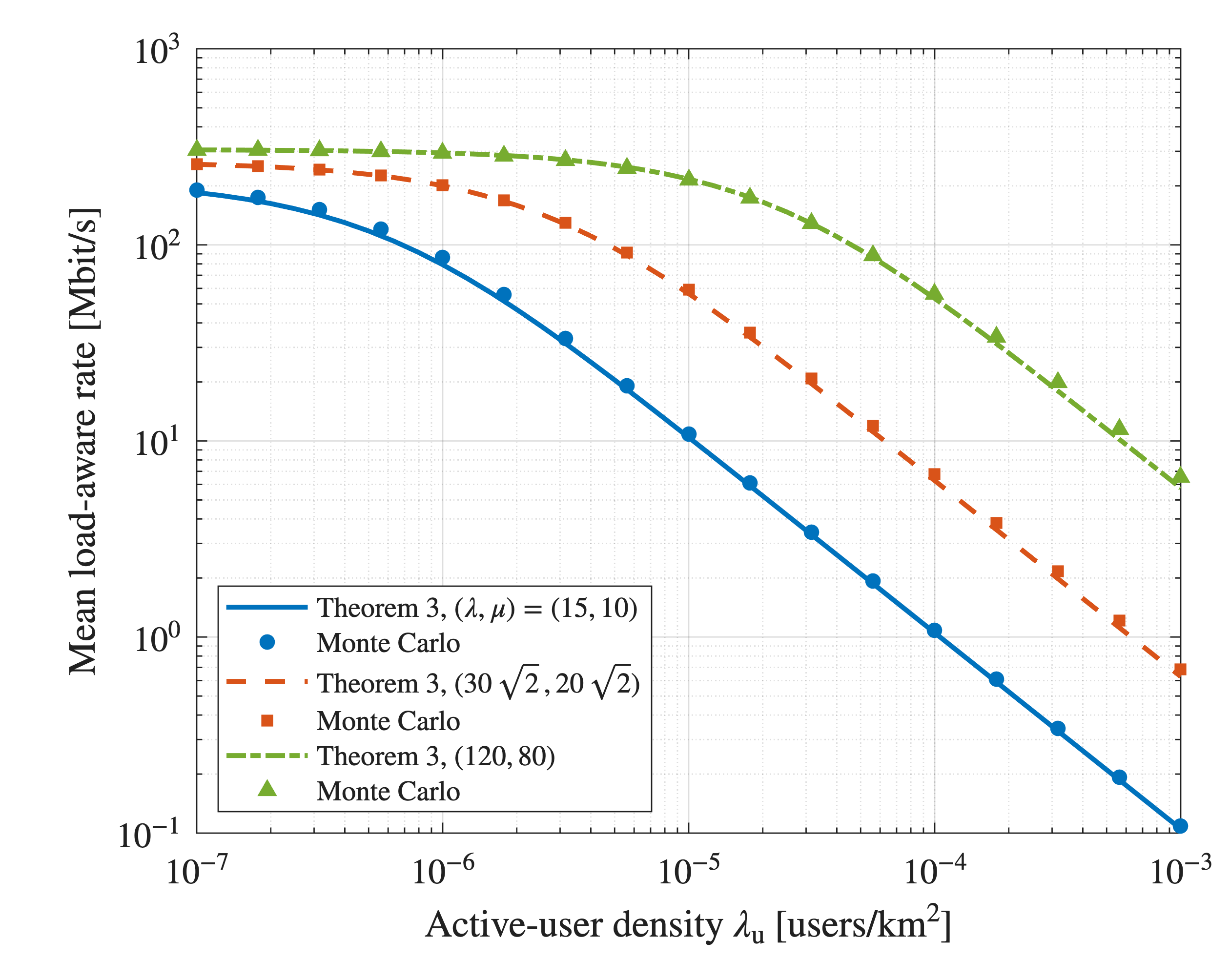}
	\caption{Mean load-aware rate versus active-user density for $(\lambda,\mu)=(15,10)$, $(30\sqrt{2},20\sqrt{2})$, and $(120,80)$ on the balanced path: Theorem~\ref{thm:rate} and Monte Carlo simulations at 17 user densities ($\gamma_{\rm f}=10^\circ$).}
	\label{fig:rate_vs_user_density}
\end{figure}

Fig.~\ref{fig:rate_vs_user_density} shows the mean load-aware rate as a function of $\lambda_{\rm u}$ for $(\lambda,\mu)=(15,10)$, $(30\sqrt{2},20\sqrt{2})$, and $(120,80)$ on the balanced path. The transition occurs near $\lambda_{\rm u}\mathbb E^0[A_{X_0}]=1$. At light load, the typical user is likely to be the only active user served by its satellite. The rate therefore approaches the link-limited value $W\bar C$. At high load, the rate decreases as $1/\lambda_{\rm u}$. The plateau heights also depend on the serving-link distribution. After coverage saturation, $\mathbb E^0[A_{X_0}]\propto1/(\lambda\mu)$ by Corollary~\ref{cor:coverage_load}, so, at fixed $k$, the load factor depends approximately on $\lambda_{\rm u}/(\lambda\mu)$. The inverse-load approximation requires $\bar L\gg1$, not merely $\bar L>1$. The load factor $\mathbb E[1/(1+N^\star)]$ corresponds to the equal-share scheduler. A channel-aware or fairness-constrained scheduler generally requires the joint distribution of loads, channel states, and allocated resources.

\begin{figure}[t]
	\centering
	\includegraphics[width=0.92\linewidth]{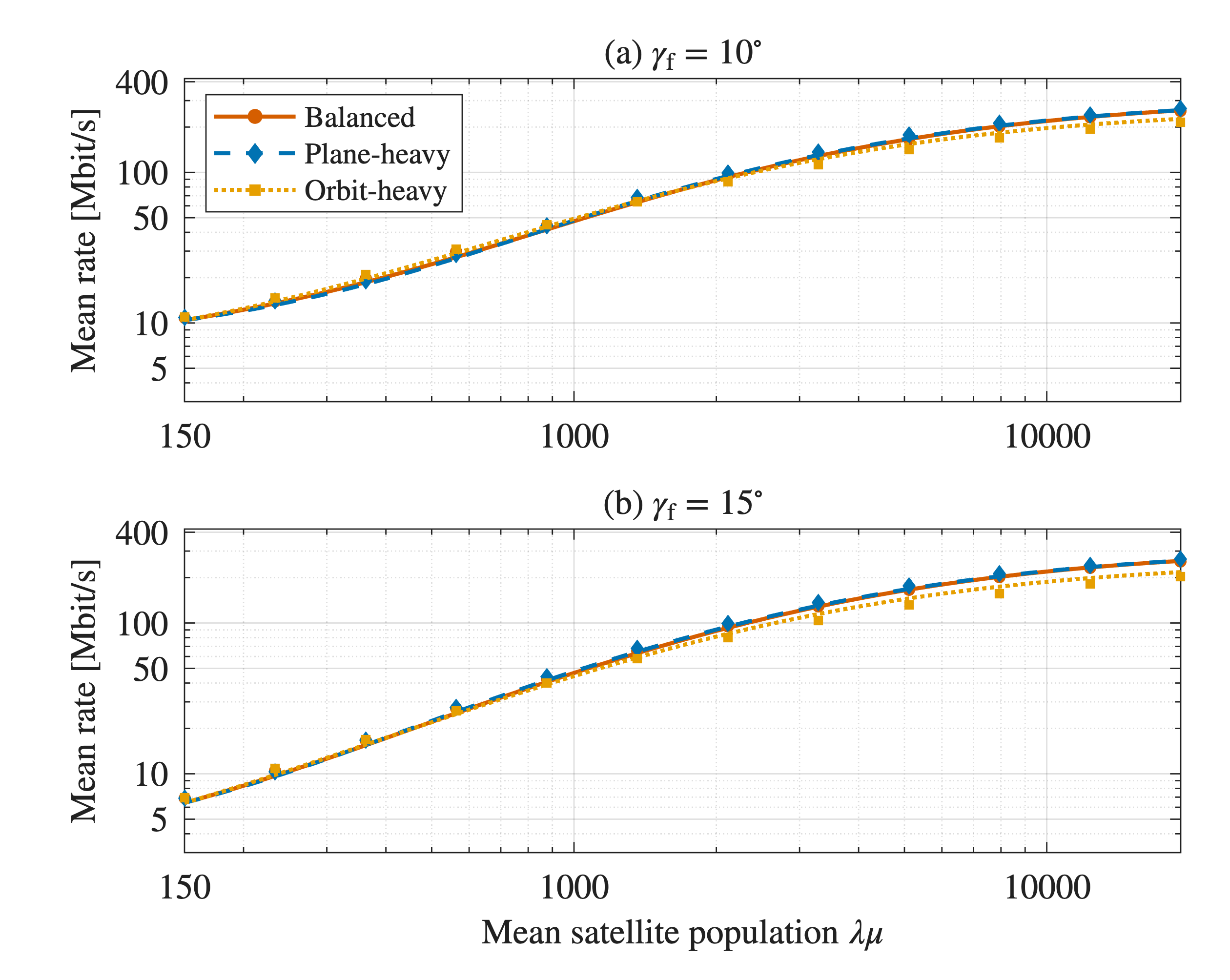}
	\caption{Mean load-aware rate under three densification paths: (a) $\gamma_{\rm f}=10^\circ$ and (b) $\gamma_{\rm f}=15^\circ$. Colors and line styles identify the paths. Curves show Theorem~\ref{thm:rate}, and markers show joint link--load Monte Carlo estimates.}
	\label{fig:densification_paths}
\end{figure}

Fig.~\ref{fig:densification_paths} shows the mean load-aware rate along the three paths for two footprint radii. The rate depends on how satellites are allocated across orbits. The balanced and plane-heavy paths are nearly indistinguishable. The Monte Carlo estimates also place the orbit-heavy path below the other two in the dense regime. The lower rate of the orbit-heavy path arises because increasing the along-orbit density does not reduce the cross-track distance to the nearest orbital plane. With a fixed number of planes, the footprint coverage approaches the ceiling $1-\exp(-\lambda\sin\gamma_{\rm f})$, since $\varrho_{\rm inter}(\gamma_{\rm f})\to\exp(-\lambda\sin\gamma_{\rm f})$ as $\mu\to\infty$. The serving-distance distribution remains less favorable as the along-orbit population grows. The two footprint radii give different rates mainly in the sparse regime. The larger footprint raises coverage but lowers the conditional rate because the additional users are served over longer slant ranges. This difference decreases for paths that approach full coverage.

Across the six plotted curves, the largest relative rate discrepancy is $11.3\%$, on the orbit-heavy path with $\gamma_{\rm f}=15^\circ$. This comparison also shows the residual error of the rate approximation away from balanced orbital allocations.

To assess how co-channel interference changes the noise-limited rate predictions, we use the following illustrative Ka-band budget: 20~GHz carrier, $W=100$~MHz, satellite effective isotropic radiated power (EIRP) of 31.3~dBW per beam, 30-dBi terminal boresight gain, and 2-dB noise figure at 290~K. At $h=550$~km these values give approximately 10-dB nadir SNR and $-122$~dBW noise power. This link budget defines a controlled sensitivity scenario and does not represent measurements of an operational constellation~\cite{kim2025coexistence,braun2019interference}.

We model the reduction in co-channel interference due to frequency reuse or beam scheduling through an independent co-channel activity probability $p$ for each other accessible satellite. With the serving-link bandwidth and power fixed, the comparison quantifies the resulting interference reduction and its effect on the load-aware rate. For a terminal pointed at its serving satellite, an interferer at off-boresight angle $\vartheta$ has relative receive gain $\max\{\exp[-\ln2(2\vartheta/5.2^\circ)^2],10^{-2.5}\}$. A unit-gain interference pattern supplies a comparison without spatial discrimination while preserving the same desired-link SNR. We use independent shadowed-Rician fading. We exclude interference from satellites outside the modeled accessibility caps and from other operators. We also exclude traffic-dependent transmitter activity.

We use common geometry and serving-link fading across the four Monte Carlo comparisons. With the directional pattern and $p=1$, the interference-aware rates are 91.9\%, 88.0\%, and 81.4\% of the noise-limited Monte Carlo rates at populations 1,200, 2,400, and 4,800. With $p=1/4$, the ratios are 96.6\%, 94.5\%, and 90.5\% at populations 2,400, 4,800, and 9,600. These comparisons quantify the error of ignoring interference for the stated model. They do not establish a universal density or mean interference-to-noise ratio (INR) threshold. At population 19,200 the $p=1/4$ rate is still increasing within the tested range, so the data do not establish saturation for that case.

\subsection{Comparison with Walker Constellations}
\label{subsec:walker}

The Cox point process is a statistical abstraction of orbit-constrained deployments. We compare it with idealized Walker constellations under the same common-altitude, footprint, association, and traffic assumptions. A Walker delta constellation has $P$ planes with ascending nodes spaced by $2\pi/P$ and $S$ satellites per plane with phases spaced by $2\pi/S$. Adjacent planes have phase offset $2\pi/(PS)$, corresponding to Walker phasing parameter one. For a single shell, we average over orbital epochs using a uniform common phase offset. We average over longitude using an independent common offset of the ascending nodes.

We use a $53^\circ$ single-inclination family and a five-component multi-inclination Starlink-type configuration with $(P,S,i)=(72,22,53^\circ)$, $(72,22,53.2^\circ)$, $(36,20,70^\circ)$, $(6,58,97.6^\circ)$, and $(4,43,97.6^\circ)$. Its total population is 4,408. All components have the same altitude of 550~km. We average over independently sampled node and phase offsets of the components. This ensemble of relative alignments provides a common-altitude geometric benchmark with the stated plane counts and inclinations.

The satellite density of a Walker shell varies with latitude. Define $i_{\max}=\arcsin(|\sin i|)\in[0,\pi/2]$, the maximum absolute latitude, including retrograde orbits. For $|\varphi|<i_{\max}$, a uniformly distributed argument of latitude yields the satellite density $PS/(2\pi^2R_{\rm s}^2\sqrt{\sin^2i-\sin^2\varphi})$ and orbit-length density $P/(\pi R_{\rm s}\sqrt{\sin^2i-\sin^2\varphi})$. Matching these with the isotropic Cox densities gives
\begin{align}
\lambda_{\rm eff}(\varphi)&=\sum_{j:|\varphi|<i_{\max,j}}\frac{2P_j}{\pi\sqrt{\sin^2i_j-\sin^2\varphi}},\nonumber\\
\mu_{\rm eff}(\varphi)&=\frac{\sum_j\lambda_j(\varphi)S_j}{\lambda_{\rm eff}(\varphi)},
\label{eq:local_matching}
\end{align}
where $\lambda_j$ is the corresponding summand and is zero outside its latitude range. The mapping is defined where $\lambda_{\rm eff}>0$ and away from the component maximum latitudes. For a single shell, $\mu_{\rm eff}=S$. For the single-shell population sweep below, we use the latitude at which the local Walker satellite density equals the mean density on the shell, $PS/(4\pi R_{\rm s}^2)$. If $|\sin i|\geq2/\pi$, this latitude is $\varphi_{\rm eq}=\arcsin\sqrt{\sin^2i-4/\pi^2}$, which is $28.8^\circ$ for $i=53^\circ$.

At each latitude, we evaluate the Cox mean area and perimeter of Theorems~\ref{thm:alpha} and~\ref{thm:beta} at the locally matched pair $(\lambda_{\rm eff}(\varphi),\mu_{\rm eff}(\varphi))$. Thus the parameters vary with latitude even though each local approximation uses the isotropic Cox model. The matching uses only the orbit and satellite densities, without fitting either service-region measure. It provides a tractable approximation to the Walker means at that latitude.

\begin{figure}[t]
\centering
\includegraphics[width=0.92\linewidth]{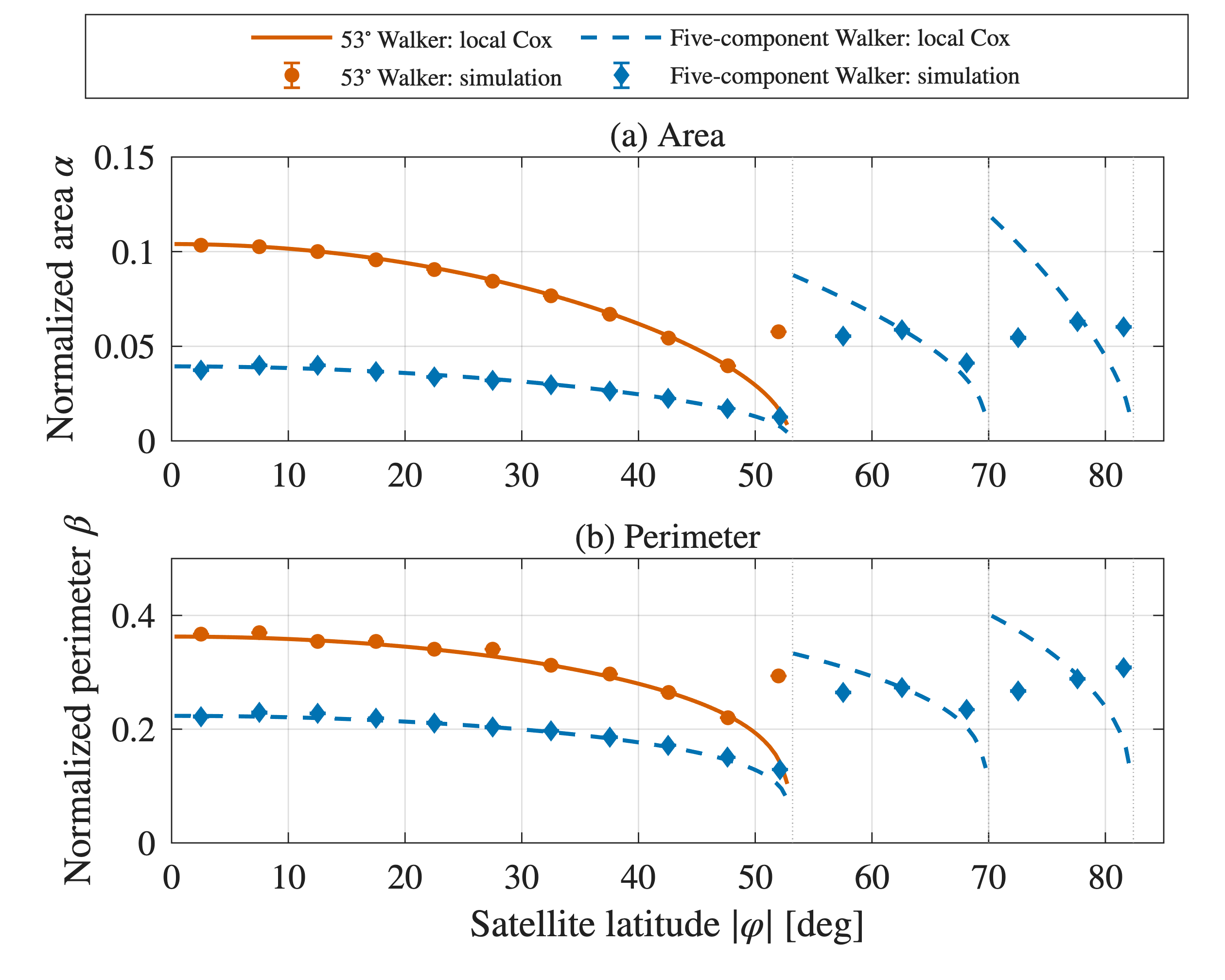}
\caption{Mean normalized (a) area and (b) perimeter versus satellite latitude for the $53^\circ$ Walker shell with $(P,S)=(72,22)$ and the five-component benchmark ($\gamma_{\rm f}=10^\circ$). Curves show the predictions of Theorems~\ref{thm:alpha} and~\ref{thm:beta} evaluated at the latitude-dependent parameters in \eqref{eq:local_matching}. Markers show $5^\circ$ bin means at their mean satellite latitudes. Error bars show $1.96$ standard errors across independent alignments and are mostly smaller than the markers. Vertical guides mark component maximum latitudes at $53.2^\circ$, $70^\circ$, and $82.4^\circ$; the $53^\circ$ limit lies next to the first guide.}
\label{fig:walker_area_perimeter_latitude}
\end{figure}

We evaluate every satellite's footprint-constrained cell by radial boundary integration and pool the area and perimeter by satellite latitude. The samples contain 253 independent alignments of the $53^\circ$ configuration and 91 alignments of the five-component configuration. We calculate standard errors across alignments to account for dependence between cells in the same constellation.

We compare latitude-bin widths of $1^\circ$, $2.5^\circ$, $5^\circ$, and $10^\circ$ using the same cells. Narrow bins resolve oscillations associated with the regular Walker placement. The $10^\circ$ bins average across a wider range of local densities and obscure changes near component maximum latitudes. We retain $5^\circ$ bins as a compromise between local resolution and averaging over this regular structure. For the five-component configuration, the median relative standard error of the mean area is $1.19\%$, $0.81\%$, $0.29\%$, and $0.13\%$, respectively. For the perimeter, the corresponding values are $0.50\%$, $0.33\%$, $0.17\%$, and $0.12\%$. We place each marker at the mean satellite latitude of its bin and evaluate the local Cox prediction there when reporting relative differences.

Fig.~\ref{fig:walker_area_perimeter_latitude} displays both mean area and mean perimeter. In the ten $5^\circ$ bins from $0^\circ$ to $50^\circ$, the five-component Walker means differ from the local Cox predictions by at most $5.4\%$ for area and $3.6\%$ for perimeter. For the $53^\circ$ configuration, the corresponding differences are below $1.9\%$ and $3.7\%$. These comparisons quantify the accuracy of the locally evaluated Cox means at the stated latitude-averaging scale. Finer bins reveal the additional variation produced by regular satellite spacing.

Larger differences occur near and immediately above component maximum latitudes. For the $53^\circ$ configuration, the final $50^\circ$--$55^\circ$ bin contains satellites only up to $53^\circ$. Its mean normalized area is $0.0577$, compared with $0.0171$ from the locally matched Cox approximation. Thus the close agreement below $50^\circ$ does not extend to this final bin. At mean latitude $57.5^\circ$, the five-component Walker area and perimeter are 0.0554 and 0.2647, compared with local Cox predictions of 0.0754 and 0.3091. At $72.5^\circ$, they are 0.0545 and 0.2669, compared with 0.1042 and 0.3732. Near a component's maximum latitude, its pointwise density changes rapidly and its orbital directions concentrate near the latitude circle. Immediately above that latitude, its contribution to \eqref{eq:local_matching} vanishes even though its nearby satellites can still restrict the service region. A pointwise match of local densities therefore need not describe the neighborhood occupied by a finite service region near an inclination limit. These effects help explain the larger differences in those neighborhoods. We conjecture that superposing different inclinations helps reduce directional bias where several components overlap.

\begin{figure}[t]
\centering
\includegraphics[width=0.92\linewidth]{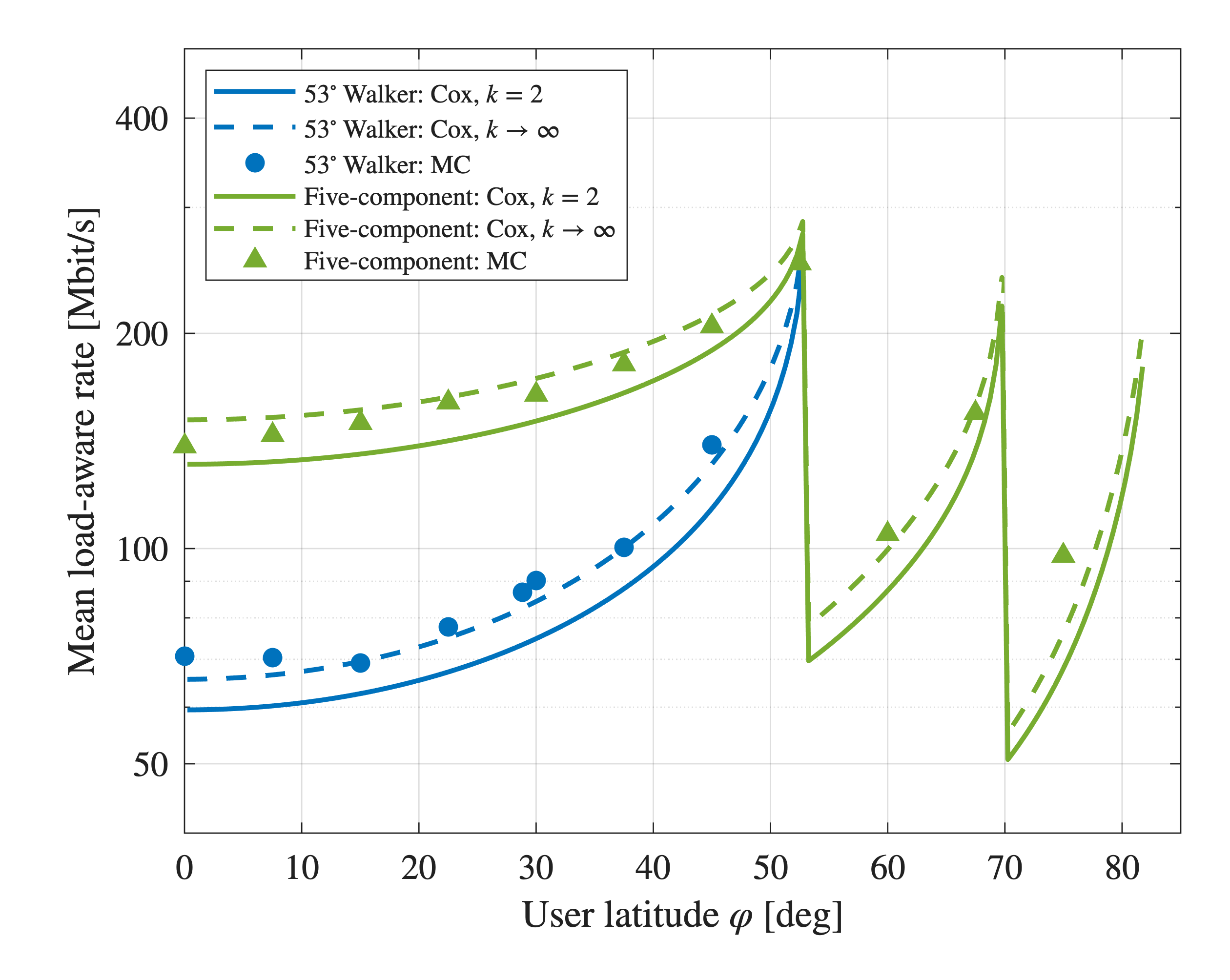}
\caption{Walker load-aware rate versus user latitude and local Cox approximations with $k=2$ and $k\to\infty$ ($\gamma_{\rm f}=10^\circ$, $\lambda_{\rm u}=10^{-5}$ users/km$^2$). Markers show Walker Monte Carlo estimates with 95\% confidence intervals. The analytical curves are comparisons, not bounds on the Walker rate.}
\label{fig:walker_rate_vs_latitude}
\end{figure}

For user-centric comparisons, we average over independently sampled alignments at each user latitude and compute each service-region area $A^\star$ by radial boundary integration. In Fig.~\ref{fig:walker_rate_vs_latitude}, the $53^\circ$ Walker rate can exceed both Cox curves. At the equator it is 70.68 Mbit/s, compared with 59.48 Mbit/s for $k=2$ and 65.64 Mbit/s for $k\to\infty$. The local density matching does not reproduce all features of the Walker serving-distance distribution. Adjusting the Gamma shape alone cannot correct this discrepancy.

The five-component rates lie between the two curves at several latitudes, but not universally. At $52.5^\circ$, the five-component Walker rate is 250.58 Mbit/s, compared with Cox predictions of 264.84 and 278.36 Mbit/s for $k=2$ and $k\to\infty$, respectively. At $75^\circ$, the corresponding values are 97.80, 68.01, and 76.23 Mbit/s. The nearly constant areas within a single-shell latitude bin and the larger variation in the mixture help explain the load-factor difference, but neither makes the local Cox rate exact. Both link geometry and the user-weighted area distribution require validation.

\begin{figure}[t]
\centering
\includegraphics[width=0.92\linewidth]{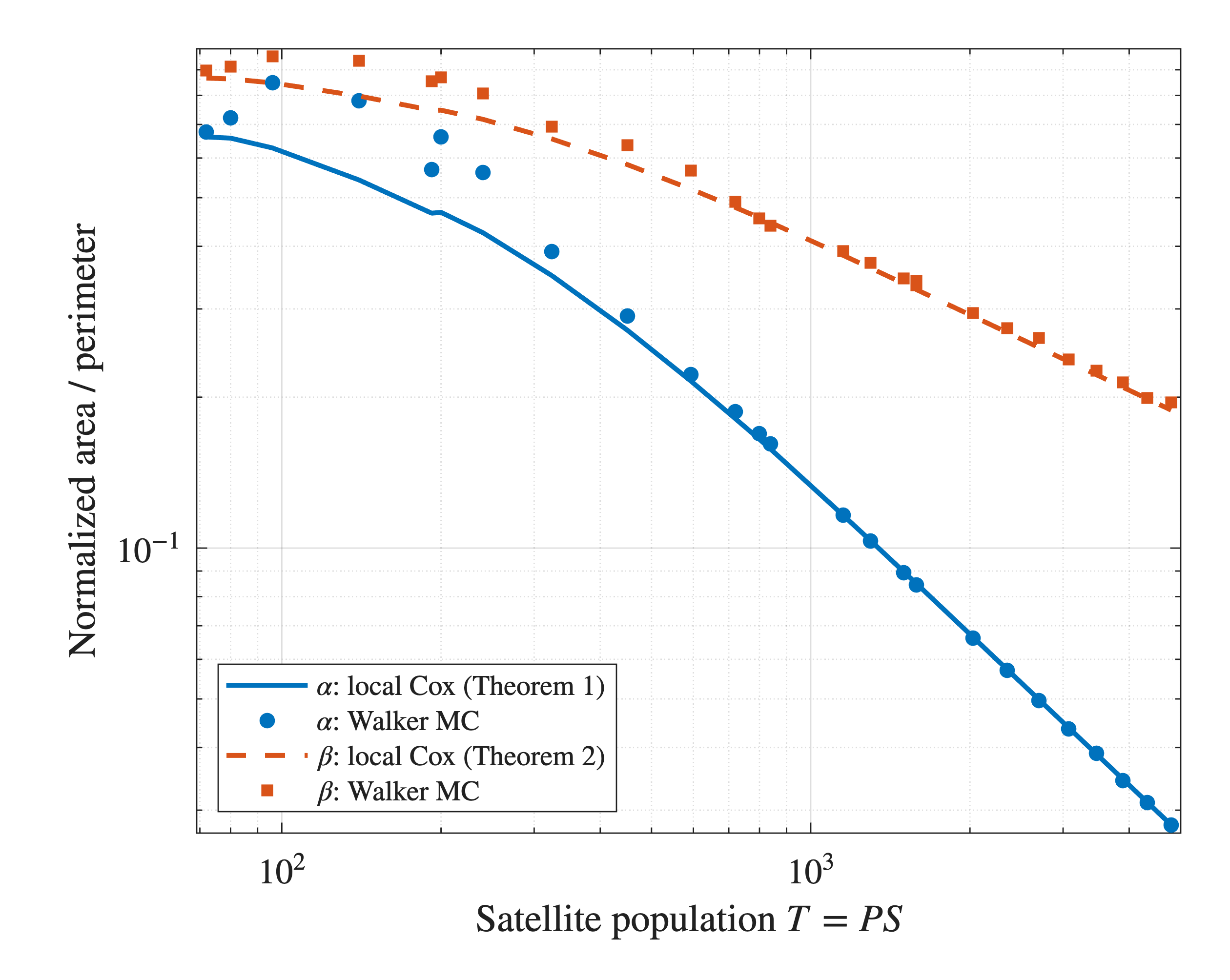}
\caption{Walker normalized area and perimeter in the $5^\circ$ bin containing $\varphi_{\rm eq}$, compared with local Cox predictions. The 26 configurations range from $(P,S)=(8,9)$ to $(120,40)$ at inclination $53^\circ$ and $\gamma_{\rm f}=10^\circ$. Markers show Monte Carlo means of independently sampled cells within this bin, with 95\% confidence intervals.}
\label{fig:walker_area_perimeter_vs_population}
\end{figure}

For the sampled single-inclination configurations with total population $T=PS\geq720$, Fig.~\ref{fig:walker_area_perimeter_vs_population} gives area errors of about 3.2\% or less and perimeter errors below 4.4\% in this bin. We sample independent satellite cells conditional on membership in the $25^\circ$--$30^\circ$ bin for each configuration. The 95\% error bars are mostly smaller than the markers. These observations do not supply a universal population threshold. Regular satellite spacing affects footprint overlap in sparse configurations. The population comparison complements the latitude-dependent comparison above.

\section{Conclusion}
\label{sec:conclusion}
We used an orbit-aware Cox process to derive the exact mean area and perimeter of footprint-constrained service regions in dense LEO networks. The coverage--load identity connects this geometry to mean satellite load. After coverage saturates, mean area and load decrease inversely with population, whereas the perimeter retains a positive limit when the mean number of orbital planes is fixed. Thus, reducing load does not imply uniform contraction of service regions. Our approximate rate analysis links these geometric effects to resource sharing: along the tested densification paths, load reduction contributes more than serving-link improvement, with diminishing returns at light load. Latitude-dependent density matching reproduces Walker mean area and perimeter over broad latitude ranges, with larger differences near component maximum latitudes. Rate predictions also depend on serving-distance and area distributions. Together, these results clarify how orbital allocation affects service boundaries, per-satellite traffic, and user rate.

Several directions remain for future research. Incorporating satellite motion would extend the present spatial analysis to the evolution of service regions and the resulting handover dynamics. Extending the model to multiple altitudes would reveal how satellites on different orbital shells jointly determine service regions and share traffic. Nonuniform user demand and load-dependent association would connect this geometry to spatial variations in traffic and available capacity.
Multi-satellite cooperation and interference-aware scheduling would further extend the analysis to service by multiple satellites, where performance depends on both their joint service regions and the allocation of resources among them.

\section*{Acknowledgment}
This work was funded by NRF RS-2024-00334240 and BK21 FOUR (Connected AI Education \& Research Program for Industry and Society Innovation, KAIST EE, No.~4120200113769).

\bibliographystyle{IEEEtran}
\bibliography{ref}

\begin{thebibliography}{10}
\providecommand{\url}[1]{#1}
\csname url@samestyle\endcsname
\providecommand{\newblock}{\relax}
\providecommand{\bibinfo}[2]{#2}
\providecommand{\BIBentrySTDinterwordspacing}{\spaceskip=0pt\relax}
\providecommand{\BIBentryALTinterwordstretchfactor}{4}
\providecommand{\BIBentryALTinterwordspacing}{\spaceskip=\fontdimen2\font plus
\BIBentryALTinterwordstretchfactor\fontdimen3\font minus
  \fontdimen4\font\relax}
\providecommand{\BIBforeignlanguage}[2]{{%
\expandafter\ifx\csname l@#1\endcsname\relax
\typeout{** WARNING: IEEEtran.bst: No hyphenation pattern has been}%
\typeout{** loaded for the language `#1'. Using the pattern for}%
\typeout{** the default language instead.}%
\else
\language=\csname l@#1\endcsname
\fi
#2}}
\providecommand{\BIBdecl}{\relax}
\BIBdecl

\bibitem{3gppNTNOverview}
{3GPP}, ``Solutions for {NR} to support non-terrestrial networks ({NTN}),'' 3rd
  Generation Partnership Project (3GPP), Tech. Rep. TR 38.821, Release 16,
  2021.

\bibitem{azari2022evolution}
M.~M. Azari, S.~Solanki, S.~Chatzinotas, O.~Kodheli, H.~Sallouha, A.~Colpaert,
  J.~F. Mendoza~Montoya, S.~Pollin, A.~Haqiqatnejad, A.~Mostaani, E.~Lagunas,
  and B.~Ottersten, ``Evolution of non-terrestrial networks from {5G} to {6G}:
  A survey,'' \emph{IEEE Commun. Surveys Tuts.}, vol.~24, no.~4, pp.
  2633--2672, 4th Quart. 2022.

\bibitem{starlinkFCC}
{Federal Communications Commission}, ``Space exploration holdings, {LLC},
  request for orbital deployment and operating authority for the {SpaceX}
  {Gen2} {NGSO} satellite system,'' FCC 22-91, Order and Authorization, Dec.
  2022.

\bibitem{kuiperFiling}
------, ``Kuiper systems {LLC}, application for authority to deploy and operate
  a {Ka}-band non-geostationary satellite orbit system,'' FCC 20-102, Order and
  Authorization, Jul. 2020.

\bibitem{simulationStudies}
I.~del Portillo, B.~G. Cameron, and E.~F. Crawley, ``A technical comparison of
  three low earth orbit satellite constellation systems to provide global
  broadband,'' \emph{Acta Astronaut.}, vol. 159, pp. 123--135, Jun. 2019.

\bibitem{ntnSharing}
{3GPP}, ``Study on management and orchestration aspects of integrated satellite
  components in a {5G} network,'' 3rd Generation Partnership Project (3GPP),
  Tech. Rep. TR 28.808, Release 17, 2021.

\bibitem{sliceNTN}
R.~Ferr{\'u}s, H.~Koumaras, O.~Sallent, G.~Agapiou, T.~Rasheed, M.-A. Kourtis,
  C.~Boustie, P.~G{\'e}lard, and T.~Ahmed, ``{SDN}/{NFV}-enabled satellite
  communications networks: Opportunities, scenarios and challenges,''
  \emph{Phys. Commun.}, vol.~18, pp. 95--112, Mar. 2016.

\bibitem{sharingTerrestrial}
J.~Kibi{\l}da, N.~J. Kaminski, and L.~A. DaSilva, ``Radio access network and
  spectrum sharing in mobile networks: A stochastic geometry perspective,''
  \emph{IEEE Trans. Wireless Commun.}, vol.~16, no.~4, pp. 2562--2575, Apr.
  2017.

\bibitem{walkerClassic}
J.~G. Walker, ``Satellite constellations,'' \emph{J. Brit. Interplanet. Soc.},
  vol.~37, pp. 559--572, 1984.

\bibitem{daley2007introduction}
D.~J. Daley and D.~Vere-Jones, \emph{An Introduction to the Theory of Point
  Processes, Volume {I}: Elementary Theory and Methods}, 2nd~ed.\hskip 1em plus
  0.5em minus 0.4em\relax Springer, 2003.

\bibitem{baccelli2010stochastic}
F.~Baccelli and B.~B{\l}aszczyszyn, ``Stochastic geometry and wireless
  networks, volume {I}: Theory,'' \emph{Found. Trends Netw.}, vol.~3, no. 3--4,
  pp. 249--449, 2009.

\bibitem{andrews2011tractable}
J.~G. Andrews, F.~Baccelli, and R.~K. Ganti, ``A tractable approach to coverage
  and rate in cellular networks,'' \emph{IEEE Trans. Commun.}, vol.~59, no.~11,
  pp. 3122--3134, Nov. 2011.

\bibitem{dhillon2012modeling}
H.~S. Dhillon, R.~K. Ganti, F.~Baccelli, and J.~G. Andrews, ``Modeling and
  analysis of {K}-tier downlink heterogeneous cellular networks,'' \emph{IEEE
  J. Sel. Areas Commun.}, vol.~30, no.~3, pp. 550--560, Apr. 2012.

\bibitem{okati2020downlink}
N.~Okati, T.~Riihonen, D.~Korpi, I.~Angervuori, and R.~Wichman, ``Downlink
  coverage and rate analysis of low earth orbit satellite constellations using
  stochastic geometry,'' \emph{IEEE Trans. Commun.}, vol.~68, no.~8, pp.
  5120--5134, Aug. 2020.

\bibitem{talgat2020stochastic}
A.~Talgat, M.~A. Kishk, and M.-S. Alouini, ``Stochastic geometry-based analysis
  of {LEO} satellite communication systems,'' \emph{IEEE Commun. Lett.},
  vol.~25, no.~8, pp. 2458--2462, Aug. 2021.

\bibitem{al2021analytic}
A.~Al-Hourani, ``An analytic approach for modeling the coverage performance of
  dense satellite networks,'' \emph{IEEE Wireless Commun. Lett.}, vol.~10,
  no.~4, pp. 897--901, Apr. 2021.

\bibitem{alhourani2021altitude}
------, ``Optimal satellite constellation altitude for maximal coverage,''
  \emph{IEEE Wireless Commun. Lett.}, vol.~10, no.~7, pp. 1444--1448, Jul.
  2021.

\bibitem{lee2024walker}
N.~Okati and T.~Riihonen, ``Nonhomogeneous stochastic geometry analysis of
  massive {LEO} communication constellations,'' \emph{IEEE Trans. Commun.},
  vol.~70, no.~3, pp. 1848--1860, Mar. 2022.

\bibitem{ppp_extensions}
D.-H. Jung, J.-G. Ryu, W.-J. Byun, and J.~Choi, ``Performance analysis of
  satellite communication system under the shadowed-{Rician} fading: A
  stochastic geometry approach,'' \emph{IEEE Trans. Commun.}, vol.~70, no.~4,
  pp. 2707--2721, Apr. 2022.

\bibitem{choi2022analytical}
C.-S. Choi and F.~Baccelli, ``A novel analytical model for {LEO} and {MEO}
  satellite networks based on {Cox} point processes,'' \emph{IEEE Trans.
  Commun.}, vol.~73, no.~4, pp. 2265--2279, Apr. 2025.

\bibitem{choi2024cox}
------, ``Cox point processes for multi-altitude {LEO} satellite networks,''
  \emph{IEEE Trans. Veh. Technol.}, vol.~73, no.~10, pp. 15\,916--15\,921, Oct.
  2024.

\bibitem{choi2024exploiting}
C.-S. Choi, ``Modeling and analysis of downlink communications in a
  heterogeneous {LEO} satellite network,'' \emph{IEEE Trans. Wireless Commun.},
  vol.~23, no.~8, pp. 8588--8602, Aug. 2024.

\bibitem{choi2026sharing}
------, ``Satellite infrastructure sharing: Orbit-structured stochastic
  geometry modeling and connectivity analysis in heterogeneous satellite
  networks,'' \emph{IEEE Trans. Wireless Commun.}, 2026, submitted for
  publication (Paper-TW-Oct-25-2810.R1).

\bibitem{choi2026uplink}
------, ``An orbit-aware {Cox} point process model for uplink communications in
  satellite networks,'' \emph{IEEE Trans. Wireless Commun.}, 2026, submitted
  for publication (Paper-TW-Jun-26-1966).

\bibitem{choi2026coordination}
------, ``Coordination-based interference management in orbit-aware {LEO}
  satellite networks: A stochastic geometry approach,'' \emph{IEEE Trans.
  Wireless Commun.}, 2026, submitted for publication (Paper-TW-Jul-26-2422).

\bibitem{choi2026walker}
C.-S. Choi and F.~Baccelli, ``Dynamical system characterization of
  heterogeneous {Walker} satellite networks: An orbit-aware stochastic geometry
  perspective,'' \emph{IEEE Trans. Wireless Commun.}, 2026, submitted for
  publication (Paper-TW-Jun-26-2296).

\bibitem{planarVoronoi}
J.~M{\o}ller, \emph{Lectures on Random {Voronoi} Tessellations}, ser. Lecture
  Notes in Statistics.\hskip 1em plus 0.5em minus 0.4em\relax Springer, 1994,
  vol.~87.

\bibitem{gammaApprox}
F.~J{\'a}rai-Szab{\'o} and Z.~N{\'e}da, ``On the size distribution of {Poisson}
  {Voronoi} cells,'' \emph{Physica A}, vol. 385, no.~2, pp. 518--526, Nov.
  2007.

\bibitem{pineda2004cell}
E.~Pineda, P.~Bruna, and D.~Crespo, ``Cell size distribution in random
  tessellations of space,'' \emph{Phys. Rev. E}, vol.~70, no.~6, p. 066119,
  Dec. 2004.

\bibitem{sphericalVoronoi}
R.~E. Miles, ``Random points, sets and tessellations on the surface of a
  sphere,'' \emph{Sankhy\={a} Ser. A}, vol.~33, no.~2, pp. 145--174, Jun. 1971.

\bibitem{kabluchko2021typical}
Z.~Kabluchko and C.~Th{\"a}le, ``The typical cell of a {Voronoi} tessellation
  on the sphere,'' \emph{Discrete Comput. Geom.}, vol.~66, no.~4, pp.
  1330--1350, 2021.

\bibitem{jeon2025design}
S.~Jeon and S.-Y. Park, ``Communication constellation design of minimum number
  of satellites with continuous coverage and inter-satellite link,'' in
  \emph{Proc. AAS/AIAA Space Flight Mech. Meeting}, 2025, paper AAS 25-254.

\bibitem{kim2025coexistence}
E.~Kim, I.~P. Roberts, and J.~G. Andrews, ``Feasibility analysis of in-band
  coexistence in dense {LEO} satellite communication systems,'' \emph{IEEE
  Trans. Wireless Commun.}, vol.~24, no.~2, pp. 1663--1677, Feb. 2025.

\bibitem{chiu2013stochastic}
S.~N. Chiu, D.~Stoyan, W.~S. Kendall, and J.~Mecke, \emph{Stochastic Geometry
  and Its Applications}, 3rd~ed.\hskip 1em plus 0.5em minus 0.4em\relax Wiley,
  2013.

\bibitem{abdi2003new}
A.~Abdi, W.~C. Lau, M.-S. Alouini, and M.~Kaveh, ``A new simple model for land
  mobile satellite channels: First- and second-order statistics,'' \emph{IEEE
  Trans. Wireless Commun.}, vol.~2, no.~3, pp. 519--528, May 2003.

\bibitem{braun2019interference}
C.~Braun, A.~M. Voicu, L.~Simi{\'c}, and P.~M{\"a}h{\"o}nen, ``Should we worry
  about interference in emerging dense {NGSO} satellite constellations?'' in
  \emph{Proc. IEEE Int. Symp. Dyn. Spectr. Access Netw. (DySPAN)}, Nov. 2019,
  pp. 1--10.

\end{thebibliography}

\end{document}